\documentclass[11pt,a4paper]{article}
\usepackage{jheppub}
\usepackage{bm,longtable,enumerate}
\usepackage{indentfirst}
\usepackage{setspace}
\usepackage{multicol}
\usepackage{multirow}
\usepackage{amsmath}
\usepackage{ulem}

\def\beq{\begin{equation}}
\def\eeq{\end{equation}}
\def\beqa{\begin{eqnarray}}
\def\eeqa{\end{eqnarray}}
\usepackage{array}

\title{Effective holographic models for QCD: Thermodynamics and viscosity coefficients}
\author[a]{Alfonso Ballon-Bayona,}
\author[b,d]{Luis A. H. Mamani,}
\author[c]{Alex S. Miranda,}
\author[d]{and Vilson T. Zanchin}
\affiliation[a]{Instituto de F\'{i}sica, Universidade
Federal do Rio de Janeiro, 
Caixa Postal 68528,  21941-972 Rio de Janeiro, Rio de Janeiro, Brazil.}
\affiliation[b]{Centro de Ci\^encias Exatas Naturais e Tecnológicas,\\ Universidade Estadual da Regi\~ao Tocantina do Maranh\~ao,\\ Rua Godofredo Viana 1300, 65901- 480, Imperatriz, MA, Brazil.}
\affiliation[c]{Laborat\'orio de Astrof\'{\i}sica Te\'orica e Observacional\ 
Departamento de Ci\^encias Exatas e Tecnol\'ogicas 
Universidade Estadual de Santa Cruz, 45650-000 Ilh\'eus, Bahia, Brazil.}
\affiliation[d]{Centro de Ci\^encias Naturais e Humanas,
Universidade Federal do ABC, 
Rua Santa Ad\'elia 166, 09210-170 Santo Andr\'e, S\~ao Paulo, Brazil.}
\emailAdd{aballonb@if.ufrj.br}
\emailAdd{luis.mamani@uemasul.edu.br}
\emailAdd{asmiranda@uesc.br}
\emailAdd{zanchin@ufabc.edu.br}

\abstract{A finite temperature extension of the effective holographic models for QCD (EHQCD),
proposed in Ref.~\cite{Ballon-Bayona:2017sxa}, is investigated in the present work. EHQCD models are characterized by two parameters, the conformal dimension of the relevant operator that deforms the CFT and the associated coupling. 
We find that black hole solutions appear at temperatures higher than some temperature $T_{min}$ and can be categorized in two classes: large and small black holes. A large black hole is thermally stable and it is therefore interpreted as the gravity dual of a non-conformal plasma. A small black hole, on the other hand, is thermally unstable. 
We show that thermodynamic quantities such as the entropy density $s$, specific heat $C_V$, and speed of sound $c_s$ are sensitive to the model parameters. We investigate perturbations of the black hole solutions and calculate the viscosity coefficients of the corresponding dual non-conformal plasma. 
For the shear viscosity, we confirm that the ratio $\eta/s$ is given by the universal result $1/4\pi$. For the bulk viscosity, the ratio $\zeta/s$ varies with the temperature, displaying a rapid growth close to $T_{min}$, and it is sensitive to the model parameters. 
We compare our results for the thermodynamic quantities with the lattice $SU(N_C)$ results and find that they are compatible as long as the coupling is fixed appropriately as a function of the conformal dimension. We also compare our results for the viscosity coefficients against the JETSCAPE results that are obtained from the analysis of experimental data on heavy ion collisions. }


\begin{document}
\maketitle

\section{Introduction}

The emergence of the Anti-de Sitter/Conformal Field Theory (AdS/CFT) correspondence \cite{Maldacena:1997re, Gubser:1998bc, Witten:1998qj} about two decades ago opened a new window to investigate the very specific properties of the quark-gluon plasma (QGP) produced for the first time at the Relativistic Heavy Ion Collider (RHIC) \cite{Adcox:2004mh, Back:2004je, Arsene:2004fa, Adams:2005dq}. 
Due to its collective (macroscopic) behavior, this plasma may be described by relativistic hydrodynamics. Moreover, the QGP phase lies in the strong coupling regime of QCD and the perturbative methods of quantum field theory are not useful. 
The AdS/CFT correspondence establishes a duality between strongly coupled conformal field theories in 4d flat space and weakly coupled gravitational or string theories in $AdS_5 \times {\cal M}$ with ${\cal M}$ a compact manifold. The most notable example of this correspondence is the duality between $\mathcal{N}=4$ Super-Yang-Mills theory in 4d flat space and Type IIB Supergravity in $AdS_5\times S^5$ (see, for instance, Refs.~\cite{Son:2002sd, Policastro:2002se, Starinets:2002br} and the references therein).

Motivated by the initial success of the AdS/CFT correspondence, phenomenological models were proposed to investigate Quantum Chromodynamics (QCD)-like theories. Since QCD is a non-conformal theory (except in the extreme ultraviolet (UV) region), a deformed CFT in which the conformal invariance has been broken by a scalar operator may be considered as a reasonable approximation for QCD.
This kind of approach can be seen as a phenomenological bottom-up model for holographic QCD. To mention a few of previous works, Gubser and Nellore \cite{Gubser:2008ny} built a holographic QCD model, where the CFT is deformed by a relevant operator, constrained by the equation of state of QCD. Viscosity coefficients were later obtained in Refs.~\cite{Gubser:2008yx, Gubser:2008sz}. 
In turn, Gursoy, Kiritsis, and Nitti \cite{Gursoy:2007er} built another holographic QCD model, where the CFT is deformed by a marginal operator. Moreover, the authors of \cite{Gursoy:2007er} found the appropriate IR behavior of the gravitational background that leads to confinement and to a linear spectrum. As a consequence, the model of \cite{Gursoy:2007er}, known as improved holographic QCD (IHQCD), leads to a glueball mass spectrum in agreement with the results of lattice QCD. 
The unconfined phase in the IHQCD model was investigated in Ref.~\cite{Gursoy:2008za} and the viscosity coefficients were obtained in \cite{Gursoy:2009kk}.

The holographic QCD models described above have specific properties that make them unique. 
Considering the ultraviolet (UV) properties of the Gubser-Nellore model \cite{Gubser:2008ny} and the confined infrared (IR) characteristics of the IHQCD model built by Gursoy, Kiritsis and Nitti \cite{Gursoy:2007er}, a simple holographic (hybrid) model was recently proposed and explored in Ref.~\cite{Ballon-Bayona:2017sxa}. 
The model parameters in \cite{Ballon-Bayona:2017sxa} are the conformal dimension $\Delta=4 - \epsilon$ of the relevant operator that deforms the CFT and the associated dimensionless coupling $\hat \phi_0$. 
The model leads to a glueball mass spectrum in good agreement with lattice QCD and it was also realized that the mass spectrum is not sensitive to the variation of the conformal anomalous dimension $\epsilon$ as long as the coupling $\hat \phi_0$ is fixed as a function of $\epsilon$ (for details, see Refs.~\cite{Ballon-Bayona:2017sxa,Ballon-Bayona:2018ddm}).

In relation to the QGP, its transport coefficients (for example, the viscosity coefficients) must be finite, even if the plasma approaches a perfect fluid. 
Thus, the knowledge of their viscosity coefficients shall provide a better understanding of the properties and the evolution of the QGP. 
One way to calculate such coefficients is to consider the low-energy limit of the real-time spectral functions and then, using the Kubo formula, obtain the desired viscosity coefficients. 
This procedure is followed by some of the research groups working on lattice $SU(N_c)$ theories (see, for instance, Ref.~\cite{Meyer:2007ic, Mages:2015rea, Pasztor:2018yae}). 
However, as it is well known, transport properties are still a very challenging problem in lattice $SU(N_c)$ theories and there are large uncertainties \cite{Bazavov:2019lgz} (for a discussion on the bulk viscosity, see for instance Ref.~\cite{Meyer:2007dy}). 
Using the holographic QCD approach, based on the AdS/CFT correspondence, one may be able to extract the viscosity coefficients of the dual plasma. 
One of the methods used is turning on a source for the desired response; for example, considering metric perturbations of the dual 5d asymptotically AdS black hole. 
Imposing incoming boundary conditions at the black hole horizon and Dirichlet boundary conditions at the AdS boundary one is able to calculate the corresponding retarded Green's functions. Then, using the Kubo formula it is possible to extract the shear and bulk viscosities of the dual plasma.

In this paper, we investigate the thermodynamic properties and viscous coefficients of a non-conformal plasma in an EHQCD approach, where the CFT 
is deformed by a relevant operator dual to a 5d scalar field (the dilaton). In this approach, the gravity dual of the non-conformal plasma is given by a 5d asymptotically AdS black hole arising from Einstein-dilaton gravity with a specific dilaton profile that matches the conformal dimension of the 4d operator in the UV and satisfies the confinement criterion in the IR. 
An important consequence of the confinement scale, present in the EHQCD model, will be the emergence of a minimum temperature $T_{min}$ for the existence of black hole solutions. 
Moreover, at fixed temperature $T>T_{min}$ there will be two types of black hole solutions, namely the large and small black holes. 
The large (small) black hole will be characterized by the entropy that increases (decreases) as a function of the temperature. 
The large black hole will be thermally stable and therefore interpreted as the gravity dual of the non-conformal plasma.

We investigate the thermodynamics and the transport properties of the non-conformal plasma and find that they are sensitive to the two relevant parameters of the EHQCD model, namely the conformal dimension $\Delta=4-\epsilon$ of the scalar operator that deforms the CFT and the corresponding dimensionless coupling $\hat \phi_0$. 
We obtain the entropy of the non-conformal plasma using the Bekenstein-Hawking formula for the black hole entropy whereas the plasma free energy will be obtained by the use of the first law of Thermodynamics. Expanding the action up to second-order we are able to extract the viscosity coefficients. 
For the shear viscosity, we rewrite the perturbations in terms of gauge-invariant variables describing the transverse and traceless sector. 
For the bulk viscosity, we change our strategy and rewrite the metric and equations of motion using the dilaton field as an independent variable. 
In such a procedure the dilaton field plays the role of the holographic coordinate, allowing us to simplify the calculations. 
For the thermodynamics, we compare our results with those found in lattice $SU(N_c)$ theories \cite{Panero:2009tv}. 
Interestingly, we find that the thermodynamic quantities can be fit for any value of the conformal anomalous dimension $\epsilon$ as long as the dimensionless coupling $\hat \phi_0$ is fixed appropriately as a function of $\epsilon$. 
This is consistent with our previous results at zero temperature \cite{Ballon-Bayona:2017sxa, Ballon-Bayona:2018ddm}, where fitting the glueball mass spectrum also led to fixing $\hat \phi_0$ as a function of $\epsilon$. 
For the viscous coefficients, we compare our results with the phenomenological constraints found by the JETSCAPE collaboration, from a model-to-data analysis of heavy ion collision experimental data \cite{Everett:2020yty}.

The present paper is organized as follows. In Sec.~\ref{Sec:5DModel} we describe the extension of the EHQCD models to finite temperature and the physical interpretation of the model parameters. 
Sec.~\ref{Sec:Thermo} deals with the thermodynamic properties of the EHQCD models, where we describe the two black hole solutions appearing at finite temperature and discuss their stability. 
In section \ref{Sec:ViscosityCoeffs} we investigate the viscosity coefficients considering the metric perturbations dual to the stress-energy tensor components associated with shear and expansion. 
Sec.~\ref{Sec:Data} is devoted to comparing our results for the thermodynamics with the results found in lattice $SU(N_C)$ theories and comparing our results for the viscous coefficients with the phenomenological constraints found by the JETSCAPE collaboration. 
We present our conclusions in Sec.~\ref{Sec:Conclusions}, and the asymptotic analysis of the EHQCD models in Appendix \ref{Sec:Asymptotics}.

\section{Effective holographic QCD models at finite temperature}
\label{Sec:5DModel}

In this section, we summarize the general properties of the effective holographic QCD models (EHQCD) and present the ansatz for the black hole solutions that allow for a description of the non-conformal plasma at finite temperature. 
In sections \ref{Sec:Thermo} and \ref{Sec:ViscosityCoeffs} we will describe the thermodynamic and transport properties of the non-conformal plasma. 
For more details on EHQCD models at zero temperature, see the discussion presented in Refs.~\cite{Ballon-Bayona:2017sxa, Ballon-Bayona:2018ddm}.

EHQCD models arise as asymptotically AdS solutions in 5d Einstein-dilaton gravitational theory. 
These 5d backgrounds are the gravity duals of 4d deformations of a CFT by a relevant scalar operator \cite{Gubser:2008ny}. 
The Lagrangian of the deformed CFT is given by
\noindent
\begin{equation} \label{eq:lagr0}
\mathcal{L}= {\cal L}_{CFT} + \phi_0\mathcal{O},
\end{equation}
\noindent
where $\phi_0$ is the source, and $\mathcal{O}$ is the scalar operator dual to the dilaton field. 
The physical motivation behind this approach is the possibility of interpreting the Lagrangian \eqref{eq:lagr0} in terms of the Yang-Mills Lagrangian in the limit of a large number of colors, $N_c$. 
The profile of the 5d dilaton field is constrained in the UV (close to the AdS boundary) and in the IR (far from the AdS boundary). 
The UV asymptotics is fixed by the relation $m^2= \Delta (\Delta-4)$ between the 5d dilaton mass and the conformal dimension of the 4d scalar operator. 
For the IR asymptotics, we follow \cite{Gursoy:2007er} where it was found a dilaton field that, far from the boundary, grows quadratically in the radial direction and leads to an asymptotically linear spectrum for the scalar and the tensor glueballs. 
Moreover, the authors of Ref.~\cite{Gursoy:2007er} showed that such a quadratic behavior also guarantees the key requirement for color confinement, namely that the warp factor in the string frame presents a global minimum \cite{Kinar:1998vq}.

The starting point in building holographic models for QCD is the five-dimensional action for Einstein-dilaton theory, 
\noindent
\begin{equation}\label{EqAction}
S= \sigma \int d^{5}x\,\sqrt{-g}\Big [
R-\frac43\,\partial^{m}\Phi\partial_{m}\Phi+V(\Phi) \Big ]
\equiv \sigma \int d^{5}x\,\sqrt{-g} \left ( R + {\cal L}_{\Phi}  \right ) ,
\end{equation}
\noindent
where $\Phi$ is the scalar (dilaton) field,  $V(\Phi)$ is  the associated potential, and $\sigma = M_p^3N_c^2$ is the effective 5D gravitational coupling. 
$M_p$ is the effective Planck mass and $N_c$ is a finite constant parameter associated with the number of colors in the four-dimensional dual field theory. 
The Einstein-dilaton equations of motion obtained from this action are
\noindent
\begin{align}
&G_{mn} =\frac{1}{2 \sigma}T_{mn},  \label{EqEinstein} \\
\frac43\,\frac{1}{\sqrt{-g}}
&\partial_{m}\left(\sqrt{-g} \, g^{mn}\partial_n\Phi\right)
+\frac12\,\frac{dV}{d\Phi} =0, \label{Eq:KG}
\end{align}
\noindent
where $G_{mn}$ stands for the Einstein tensor and $T_{mn}$ is the energy-momentum tensor defined by 
\noindent
\begin{equation}
T_{mn}= - \frac{2}{\sqrt{-g}} \frac{\delta S_M}{\delta g^{mn}} = 
\sigma \left[\frac83\,\partial_{m}\Phi\partial_n\Phi
+g_{mn} {\cal L}_{\Phi} \right],
\end{equation}
\noindent
with $S_{M} = \sigma \int d^5 x {\cal L}_{\Phi}$ being the matter action due to the scalar field. 

Let us notice that Eqs.~\eqref{EqEinstein} and \eqref{Eq:KG} are expressed in general forms and then they apply to any spacetime geometry. 
However, in order to investigate holographic QCD models at finite temperature we consider the following black hole metric 
\noindent
\begin{equation}\label{EqBHmetric}
ds^2=\frac{1}{\zeta_1(z)^2}\left(-f(z)dt^2
+\frac{1}{f(z)}dz^2+dx_idx^i\right),
\end{equation}
\noindent
where $f(z)$ is the horizon function (or blackening function), $\zeta_1(z)$ is a global (inverse) scale factor for the 5D metric related to the warp factor through the relation $\zeta_1=e^{-A_1}$, 
with $A_1$ being the warp factor \footnote{In the present work we use subscripts to distinguish the scale factor $\zeta_1$ from the bulk viscosity $\zeta$.}. 
The black hole solution is characterized by the presence of an event horizon, $z_h$, where the horizon function vanishes. 
With this, the holographic coordinate $z$ is restricted to the interval $0\leq z\leq z_h$. 
A confining thermal solution can also be investigated by setting $f(z)$ to $1$ for $0<z<\infty$. 
This is the trivial extension of the zero temperature solution dual to a confining vacuum. 
To describe the physics of a dual non-conformal plasma, in this work we focus on a black hole solution 
corresponding to a nontrivial $f(z)$ that interpolates between $f(0)=1$ at the AdS boundary and $f(z_h)=0$ at the event horizon.

After plugging \eqref{EqBHmetric} into \eqref{EqEinstein}, the equations of motion can be written in the form
\noindent
\begin{equation}\label{EqsBH}
%\begin{split}
\zeta_1''
-\frac{4}{9}\Phi'^{\,2}\,\zeta_1=\,0,\qquad
% V+3\zeta_1^5\left(\frac{\zeta_1'\,f}{\zeta_1^4}\right)'=\,0,\qquad
\zeta_1^{-5} V - \left(\left( \zeta_1^{-3}\right)' f \right)'=\,0,\qquad
% \left(\frac{f'}{\zeta_1^3}\right)'=\,0,
\left(\zeta_1^{-3} f' \right)'=\,0,
%\end{split}
\end{equation}
\noindent
where the prime indicates the total derivative with respect to $z$. The coupled equations \eqref{EqsBH} may be solved by following different approaches. 
The traditional method is to build a dilaton potential that interpolates between the UV and the IR regime and then solve the equations for the fields $\Phi$, $\zeta_1$, and $f$. This is the strategy considered in the pioneer works of \cite{Gubser:2008ny} and \cite{Gursoy:2007er}. 
As described in our previous work \cite{Ballon-Bayona:2017sxa}, an alternative strategy for solving the equations \eqref{EqsBH} is to presuppose that we know either the dilaton field $\Phi$ (models A) or the scale factor $\zeta_1$ (models B). 
One can build simple interpolations in any of those scenarios that satisfy the UV and IR constraints. 
The potential is then reconstructed by the use of the 2nd equation in \eqref{EqsBH}. 
Any of those two scenarios are usually known as the potential reconstruction method and have been successfully used in previous works, 
see for instance Refs. \cite{Li:2013oda, Yang:2014bqa, Dudal:2017max, Arefeva:2020byn}.

In the models of type A, where the dilaton profile is known, the (inverse) scale factor $\zeta_1$ is calculated by solving numerically the first equation in~\eqref{EqsBH}. 
Then, solving the last equation in~\eqref{EqsBH}, we get an expression for the function $f(z)$ in terms of the warp factor. 
The two integration constants arising in the blackening function are fixed by considering two boundary conditions. 
Namely, a condition imposed at the AdS boundary, where $f(0)=1$, while the other one is given by the regularity condition at the horizon, 
where $f(z_h)=0$. Thus, the blackening function results in
\noindent
\begin{equation}\label{EqBlackening}
f(z)=1-C_{h}\int_{0}^{z}d\tilde{z}\,\zeta_1^3(\tilde z),
\end{equation}
\noindent
where the constant $C_h$, which guarantees the regularity condition $f(z_h)=0$, is given by
\noindent
\begin{equation}\label{EqIntegConst}
C_h=\frac{1}{\int_{0}^{z_h} d\tilde z\,\zeta_1^3(\tilde z)}.
\end{equation}
\noindent
The (inverse) scale factor $\zeta_1$ is a non-negative monotonic function of $z$ and hence the constant $C_h$ is non-negative. 
This in turn gives rise to a monotonically decreasing function $f(z)$ that goes from $f(0)=1$ at the AdS boundary to $f(z_h)=0$ at the horizon. 

In this work, we describe the finite temperature extension of the EHQCD model of \cite{Ballon-Bayona:2017sxa}. 
We restrict ourselves to the model of type A1, where the dilaton profile is given by
\noindent
\begin{equation}\label{Eq:DilModelA1}
\Phi(z)=\hat\phi_0\,\left(\Lambda z\right)^{\epsilon}+
\frac{\left(\Lambda z\right)^{4-\epsilon}}{1+(\Lambda z)^{2-\epsilon}}.
\end{equation}
\noindent

As described in \cite{Ballon-Bayona:2017sxa}, in the zero temperature case the EHQCD background obtained from the dilaton profile \eqref{Eq:DilModelA1} gives rise to a glueball spectrum 
in agreement with lattice QCD results \cite{Meyer:2004gx} and the results from improved holographic QCD models \cite{Gursoy:2007er}. 
As stressed in \cite{Ballon-Bayona:2017sxa}, the glueball spectrum does not change significantly when we consider the conformal anomalous dimension, $\epsilon$, in the interval $\epsilon\in[10^{-3},10^{-1}]$. 
One of the aims of this paper is to investigate if the relevant thermodynamic variables are sensitive to the variation of the conformal anomalous dimension. 
The parameters $\hat \phi_0$ and $\Lambda$ are related to the source $\phi_0$, and vacuum expectation value $G$ by 
\noindent
\beq
\phi_0 = \hat \phi_{0}\, \Lambda^{\epsilon}, \quad \quad 
G = \Lambda^{4-\epsilon}, \,\quad\quad C=\Lambda^2. 
\label{DefParameters}
\eeq
\noindent
The parameter $C=\Lambda^2$ appears as a coefficient in the large $z$ behavior $\Phi(z) = C z^2 $ and it is related to color confinement. 
The parameter  $\Lambda$, with conformal dimension $1$, plays a role similar to $\Lambda_{QCD}$ and fixes the units of the EHQCD model. 
The parameter $\hat \phi_0$ is the dimensionless version of the source (coupling) $\phi_0$.

Thus, the relevant parameters of the model are $\hat \phi_0$, $\epsilon$, and $\Lambda$. 
Note that the parameter $\Lambda$ controls the breaking of conformal symmetry, in the limit $\Lambda \to 0$ the dilaton field vanishes, and conformal symmetry is restored. 
In this paper, we will work most of the time with dimensionless quantities so there will be no need of fixing $\Lambda$. 
In other words, we work in units where $\Lambda=1$. 

It is worth mentioning that, although the parameters $\epsilon$ and $\hat\phi_0$ are in general independent, as shown in \cite{Ballon-Bayona:2017sxa}, 
at zero temperature the glueball spectrum is a physical constraint to fit $\hat \phi$ as a function of $\epsilon$. 
A similar scenario occurs in the finite temperature extension considered in this work, where initially we take $\epsilon$ and $\hat\phi_0$ as independent parameters and will 
find that the thermodynamic properties become a physical constraint that fixes $\hat \phi_0$ as a function of $\epsilon$. 
Finally, it is also worth pointing out that in the limit $\hat\phi_0 \to 0$ a massless mode arises in the scalar sector \cite{Mamani:2019mgu}.

\section{Thermodynamics of the non-conformal plasma}
\label{Sec:Thermo}

In this section we investigate the thermodynamic properties of the black hole solutions arising in our EHQCD model, 
which will be interpreted in terms of the 4d non-conformal plasma. 

\subsection{Temperature and entropy}
\label{SubSec:Temp}

After finding the blackening function \eqref{EqBlackening}, it is 
possible to obtain the Hawking temperature of the black hole using the well-known relation
\begin{equation}\label{EqTemperature}
T=-\frac{f'(z_h)}{4\pi}=\frac{C_h \zeta_1^3(z_h)}{4\pi}.
\end{equation}
\noindent
According to the AdS/CFT correspondence, the black hole temperature obtained from the gravitational solution should be equal to the temperature of the dual quantum field theory. 
In general, the temperature is determined numerically after setting the model parameters. 
In Appendix \ref{Sec:Asymptotics} we write the asymptotic behavior of the temperature for small $z_h$ (close to the boundary), 
see Eq.~\eqref{EqTUV}, and for large $z_h$ (far from the boundary), see Eq.~\eqref{eq:TIR}. 
From the asymptotic behavior, we conclude that the temperature is non-monotonic and hence must has a minimum in the intermediate region $T=T_{min}$ at $z_h={z_h}_c$. 
In the following, we consider the case where the critical temperature for deconfinement in the non-conformal plasma is $T_c=T_{min}$. 
We shall see below that there are two black hole solutions for each temperature $T > T_c$, although, as we will show later, one of them is unstable. 

In order to investigate the effects of finite temperature in the dual field theory, we need to calculate some of the thermodynamic variables, 
such as entropy, free energy, and specific heat. In this paper, we follow a method commonly used in the literature (see, for instance, \cite{Gursoy:2008za, Gubser:2008yx}, 
for an earlier study see \cite{Andreev:2007zv}) in which the entropy density of the 5d black hole, 
obtained from the Bekenstein-Hawking area formula, is identified with the entropy density of the dual field theory. 
From the knowledge of the entropy density and the use of the first law of Thermodynamics, 
one then obtains the free energy density which, in turn, allows us to calculate the other thermodynamic quantities. 

To obtain the entropy density, we consider the transverse components of the metric \eqref{EqBHmetric} and calculate the event-horizon area through
\noindent
\begin{equation}
\mathcal{A}=\frac{1}{\zeta_1^3(z_h)}
\int d^3 x  =\frac{V_3}{\zeta_1^3(z_h)},
\end{equation}
\noindent
where $V_3$ represents the volume of the transverse Euclidean space. 
We may now calculate the entropy, using the Bekenstein-Hawking formula,
\noindent
\begin{equation}\label{Eq:Entropy}
S=\frac{\mathcal{A}}{4 G_5}=\frac{V_3}{4G_5\,\zeta_1^3(z_h)},
\end{equation}
\noindent
where $G_5$ is the five-dimensional (5D) Newton constant, which is given in terms of the model parameters by $G_5=1/(16\pi \sigma)$. 
Finally, by substituting \eqref{EqTemperature} into \eqref{Eq:Entropy} we find the thermodynamic relation for the entropy density $s=S/V_3$, 
\noindent
\begin{equation}\label{Eq:EntropyDensity}
s=\frac{4\pi\,\sigma}{\zeta_1^{3}(z_h)}=\frac{
N_{\scriptscriptstyle{c}}^2}{45\pi^2} \frac{C_h}{T},
\end{equation}
\noindent
where the relation $M_p^3=1/(45\pi^2)$ has been used in the last equality \cite{Gursoy:2008za}.

\subsection{The free energy and the trace anomaly}
\label{SubSec:FreeEn}

So far we have obtained the temperature and entropy density. Now we use the first law of Thermodynamics \footnote{We consider a system where the volume $V_3$ is fixed and hence the work is zero.},
\noindent
\begin{equation} \label{eq:freeF}
dF = -s \, dT , 
\end{equation}
\noindent
in order to calculate the Helmholtz free energy density $F$. We may integrate this equation to calculate the free energy density of the black hole. 
Note that the entropy density and temperature are explicit functions of $z_h$ and therefore we may rewrite the integration of \eqref{eq:freeF} by using $z_h$ as the independent variable,
\noindent
\begin{equation}
\int_{F_{a}}^{F} d{F}=-\int_{z_{h_a}}^{z_h} s(\tilde{z}_h)\left(\frac{dT(\tilde{z}_h)}{d\tilde{z}_h}\right)d\tilde{z}_h,
\end{equation}
\noindent
where $z_{h_a}$ is an arbitrary initial value for the horizon radius with the corresponding to the free energy $F_a$.

One might consider the prescription where the free energy density $F_{a}$ is set to zero in the limit $z_{h_a} \to \infty$, as done in Ref.~\cite{Gursoy:2008za}.  
Since in this limit the horizon function $f(z)$ tends to unity, such a prescription is equivalent to setting the free energy density of the (confining) thermal solution to zero. 
In this work, however, we consider an alternative scheme where we integrate from a finite value of $z_h$, namely, from $z_{h_a}\equiv z_{h_c}$, so that the integral representation of the free energy becomes
\noindent
\begin{equation}\label{Eq:FreeEnergy}
{F}=-\int_{z_{h_c}}^{z_h} s(\tilde{z}_h)\left(\frac{dT(\tilde{z}_h)}{d\tilde{z}_h}\right)d\tilde{z}_h.
\end{equation}
Furthermore, set the free energy at the critical temperature $T(z_{h_c})=T_c$ to zero. 
This assumption guarantees that the free energy of the physical black hole solution, namely the large black hole with $z_h < z_{h_c}$, satisfies the physical constraint $F \leq 0$. 
This constraint results because in that regime the entropy density and the temperature are monotonically decreasing functions of $z_h$. 
Interestingly, the free energy density of the small black hole solution, corresponding to the regime $z_h > z_{h_c}$, 
will also satisfy the constraint $F \leq 0$ because an inversion of the monotonic behavior of the temperature will compensate the inversion of integration limits. 
In particular, the free energy density of the small black hole solution will not vanish in the limit $z_h \to \infty$, 
where the small black hole solutions reduce to the confining thermal solutions \footnote{For a discussion of the relation between the small black hole solution and confinement see \cite{Dudal:2017max}.}.
 
After integrating by parts, we can  rewrite \eqref{Eq:FreeEnergy} in the form
\noindent
\begin{equation}
{{F}}+s\,T=
%s(z_{h_c})\,T(z_{h_c})+
\int_{z_{h_c}}^{z_h} T(\tilde{z}_h)\left(\frac{ds(\tilde{z}_h)}{d\tilde{z}_h}\right)d\tilde{z}_h.
\end{equation}
\noindent
Note that the l.h.s is precisely the definition of the energy density, which is given by
\noindent
\begin{equation}
\rho={F}+s\,T.
\end{equation}
\noindent
Therefore, the energy density may be calculated directly from the formula
\noindent
\begin{equation}
{\rho}=
%s(z_{h_c})\,T(z_{h_c})+
\int_{z_{h_c}}^{z_h} T(\tilde{z}_h)\left(\frac{ds(\tilde{z}_h)}{d\tilde{z}_h}\right)d\tilde{z}_h.
\end{equation}
\noindent

We will be interested in investigating finite temperature effects on the trace anomaly of the energy-momentum tensor which is given by $\left\langle T^{\mu}_{\mu}\right\rangle=\rho-3p$, 
where the pressure is $p=-F$. The trace anomaly can be written in the integral form
\noindent
\begin{equation}
\begin{split}
\left\langle T_{\mu}^{\mu}\right\rangle=&\int_{0}^{z_h} T^4\frac{d}{d \tilde z_h}\left(\frac{s}{T^3}\right)d\tilde{z}_h.
\end{split}
\end{equation}
\noindent
In arriving to this equation we have considered the difference  $\left\langle T_{\mu}^{\mu}\right\rangle (z_h)$ -  $\left\langle T_{\mu}^{\mu}\right\rangle(z_h^*)$ 
and used the property that the trace anomaly vanishes in the limit $z_h^* \to 0$, assuring that conformal symmetry is restored. 

The behavior of the trace anomaly at zero temperature was investigated in Ref.~\cite{Ballon-Bayona:2017sxa}. 
In the next subsection, we are going to investigate the effects of the temperature on such an important quantity. 
It is worth mentioning that, even though in some asymptotic regimes the analytic solutions can be obtained, as described in Appendix \ref{Sec:Asymptotics}, 
in the following analysis the thermodynamic variables shall be determined numerically,

\subsection{Numerical results}
\label{SubSec:NumThermod}

Let us turn the attention to our numerical results for the thermodynamic variables. 
We will describe the thermodynamic variables and coordinates in units where $\Lambda=1$. 
For different choices of $\Lambda$, for example, the values considered in \cite{Ballon-Bayona:2017sxa}, 
the thermodynamic variables and coordinates must be properly rescaled. For example, 
the temperature $T$ in units where $\Lambda=1$ is actually the dimensionless ratio $T/\Lambda$ for a different choice of $\Lambda$.

In the following analysis, we consider the parameters $\epsilon$ and $\hat\phi_0$ as independent of each other. 
One way of describing the evolution of the thermodynamics quantities with the model parameters is fixing one of the parameters and varying the other. 
Consider, for example, setting the conformal anomalous dimension to $\epsilon=0.07$ (the value used in Ref.~\cite{Gubser:2008yx}) and varying the dimensionless coupling $\hat \phi_0$ in the region  $0\leq\hat\phi_0\leq 6$.

\begin{figure}[ht]
\centering
\includegraphics[width=7cm]{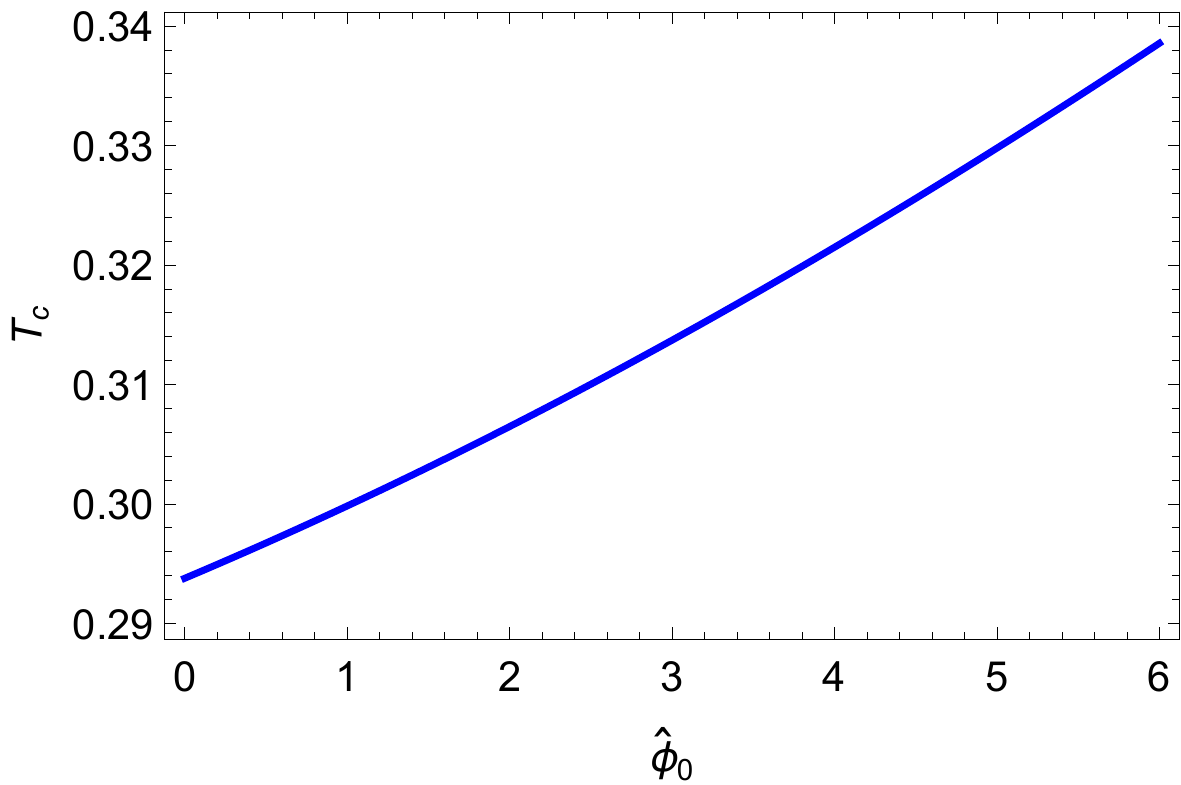}\hfill
\includegraphics[width=7cm]{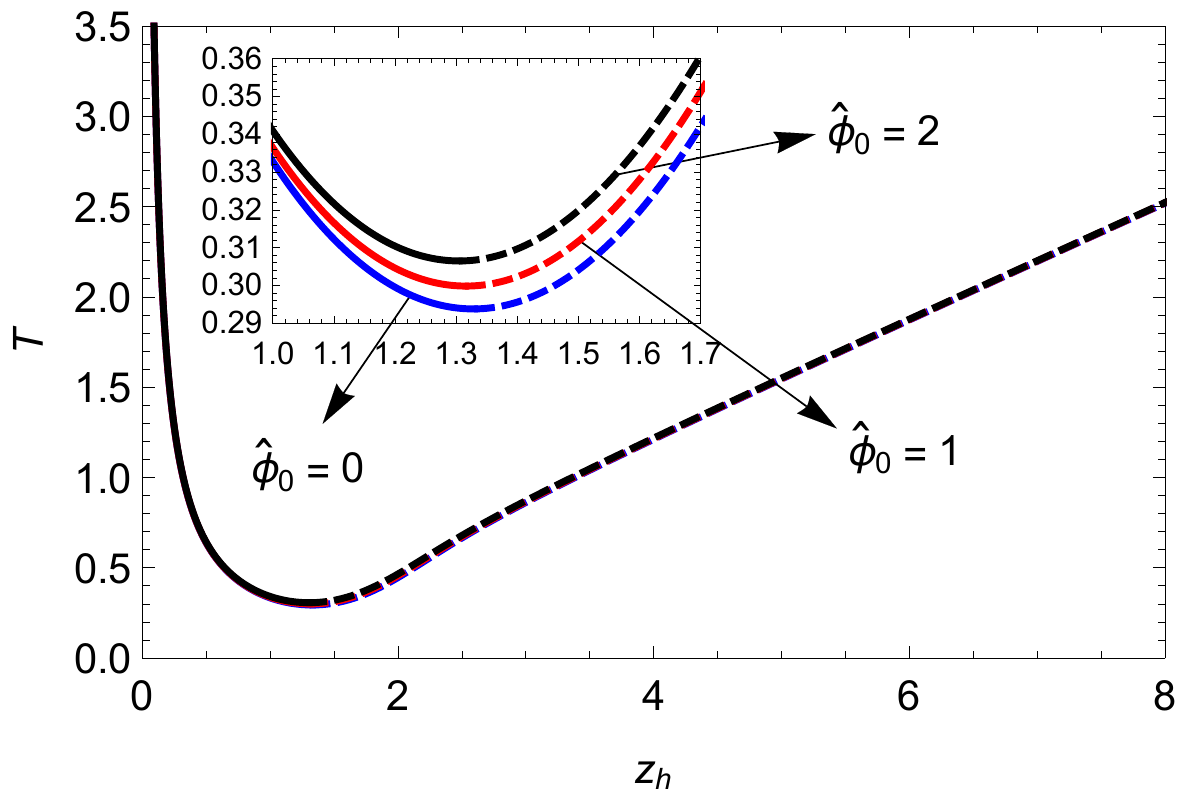}
\caption{
Left: The critical temperature $T_c$, in units where $\Lambda=1$, as a function of $\hat\phi_0$ for $\epsilon=0.07$. Right: The temperature $T$ as a function of $z_h$, both in units where $\Lambda=1$,  for three different values of $\hat\phi_0$ and $\epsilon=0.07$.
}
\label{Fig1:Phi0TFAP}
\end{figure}

The minimum temperature for the existence of black hole solutions, $T_c$, is obtained by solving the equation $\partial_{z_h}T=0$. 
On the left panel of Fig.~\ref{Fig1:Phi0TFAP}, we show that $T_c$ monotonically increases with $\hat\phi_0$. 
A plot of the temperature as a function of $z_h$, both in units where $\Lambda=1$, for the few particular values $\hat\phi_0=\{0,1,2\}$ is displayed in the right panel of Fig.~\ref{Fig1:Phi0TFAP}. 
The two branches, one of them corresponding to the large black hole regime (solid lines), and the other corresponding to the small black hole regime (dashed lines) are clearly seen. 
The inset figure displays a zoom of the region where the temperature reaches a minimum. 
We conclude that the results are sensitive to the value of $\hat\phi_0$, namely the critical temperature $T_{c}$ increases with the increasing of $\hat\phi_0$.

\begin{figure}[ht]
\centering
\includegraphics[width=7cm]{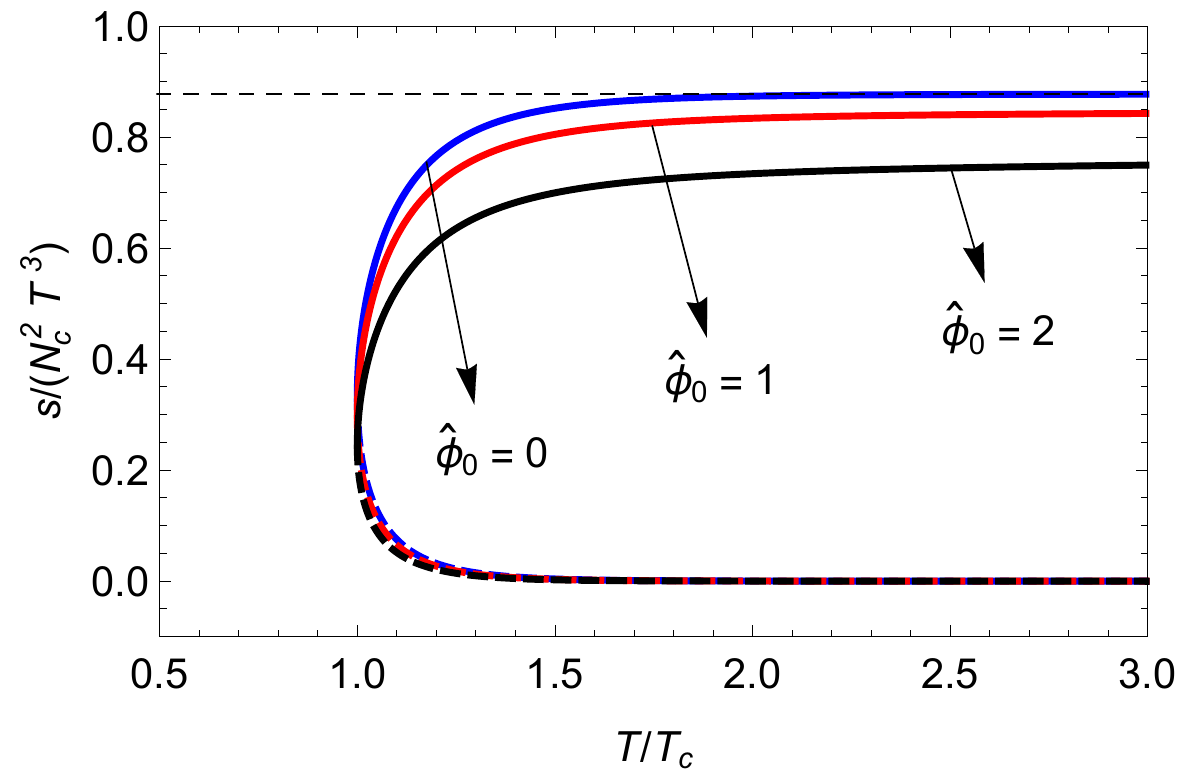}\\
\caption{
The dimensionless ratio $s/(N_{c}^{2}T^3)$ for the entropy density as a function of the normalized temperature $T/T_c$ for three different values of $\hat\phi_0$ and $\epsilon=0.07$. Solid (dashed) lines indicate   the large (small) black hole regime. The thin dashed (horizontal) line represents the value $s/(N_c^2 T^3)=4\pi^2/45$.
}
\label{Fig1:zhTFAP}
\end{figure}

The numerical results for the dimensionless entropy  $s/(N_{c}^{2}T^3)$ as a function of the dimensionless temperature $T/T_c$ are displayed in Fig.~\ref{Fig1:zhTFAP}. 
We observe that the dimensionless entropy is even more sensitive to the value of the parameter $\hat\phi_0$, it decreases with the increasing of $\hat\phi_0$. 
As before, continuous lines represent the large black hole regime, while dashed lines correspond to the small black hole regime. 
Note that the case $\hat\phi_0=0$ reaches the conformal limit, indicated by the dashed thin horizontal line, faster than the other cases. 

\begin{figure}[htb!]
\centering
\includegraphics[width=7cm]{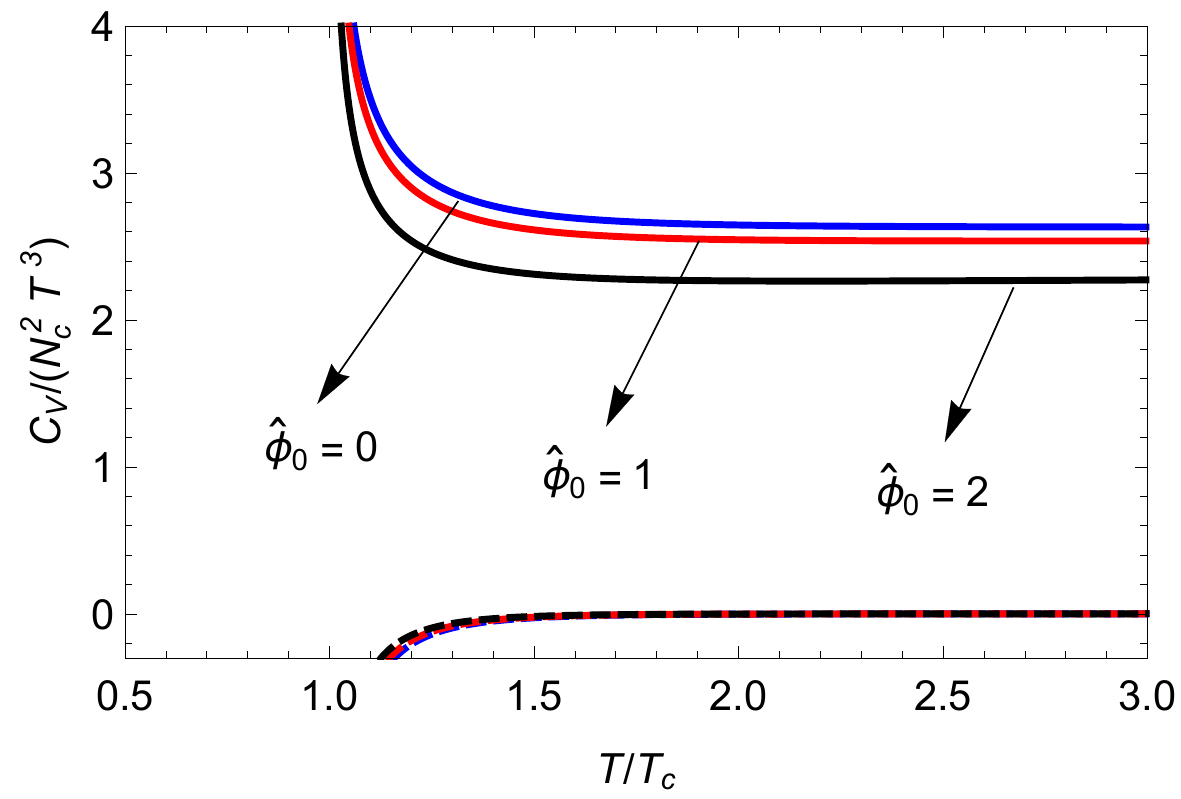}\hfill
\includegraphics[width=7cm]{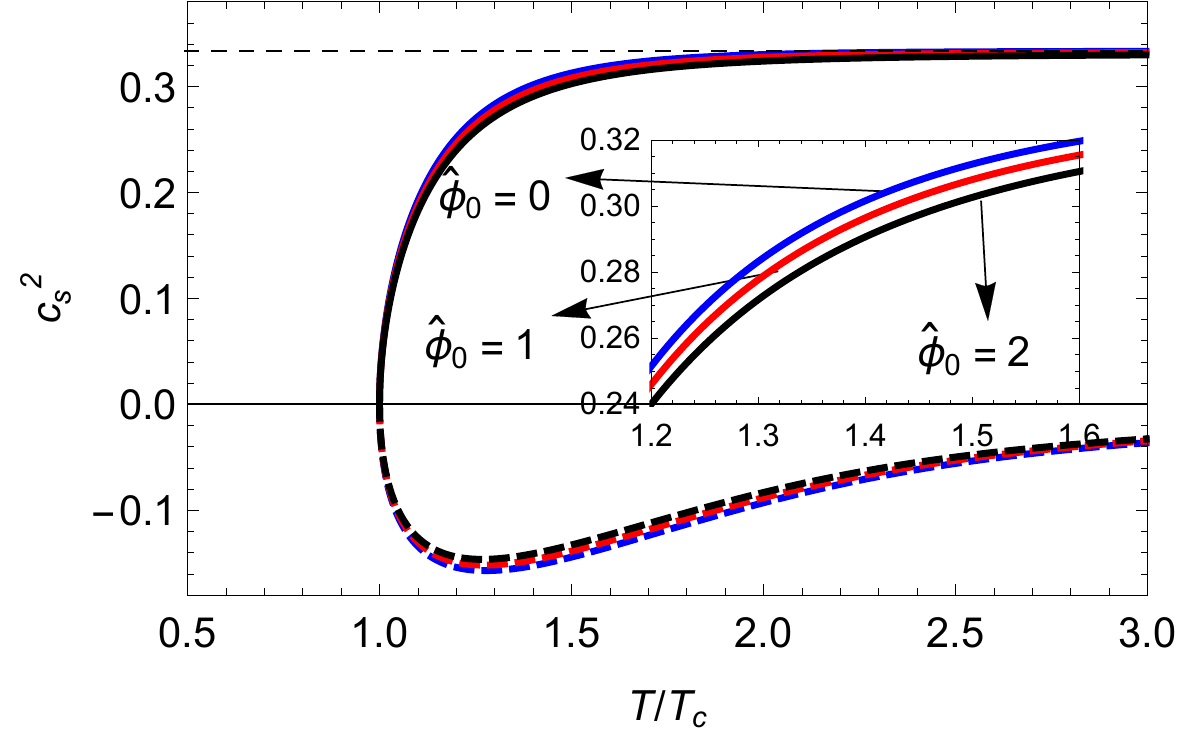}
\caption{
Left: The specific heat $C_V/(N_{c}^{2}T^3)$ as a function of $T/T_c$ for three different values of $\hat\phi_0$, as indicated, and $\epsilon=0.07$. Right: The speed of sound $c_s^2$ as a function of $T/T_c$ for three different values of $\hat\phi_0$, as indicated, and $\epsilon=0.07$. Solid (dashed) lines represent   the large (small) black holes regime. The thin dashed horizontal line in the right panel represents the value $c_s^2=1/3$.
}
\label{Fig1:TCVFAP}
\end{figure}

So far, we have two black hole regimes. In order to discover which regime is physically relevant, we must investigate the thermodynamic stability of the corresponding solutions. 
A standard method for investigating the stability of a thermodynamic system is to calculate the corresponding specific heat, defined by the relation
\noindent
\begin{equation}
C_V=T\left(\frac{\partial s}{\partial T}\right)_{V}.
\end{equation}
Stable systems are characterized by a positive $C_V$. On the left panel of Fig.~\ref{Fig1:TCVFAP} we display the results of the dimensionless specific heat, $C_V/(N_c^2 T^3)$, as a function of the dimensionless temperature $T/T_c$ for different values of $\hat\phi_0$. Solid lines represent the large black hole solutions, 
for which the specific heat is always positive (stable regime) whereas dashed lines represent the small black hole solutions, for which the specific heat is always negative (unstable regime). The figure also shows the dependence of the specific heat on the parameter $\hat\phi_0$. 
For the large (small) black hole, increasing the value of $\hat\phi_0$ leads to a decrease (increase) of the value of $C_V/(N_c^2T^3)$ for a fixed temperature.

Another relevant thermodynamic variable is the speed of sound, which may be calculated using the formula
\noindent
\begin{equation} \label{Eq:cs2}
c_s^2=\frac{\partial \ln T}{\partial \ln s}.
\end{equation}
\noindent
Our numerical results for $c_s^2$ as a function of the temperature are displayed on the right panel of Fig.~\ref{Fig1:TCVFAP} for different values of $\hat\phi_0$, as indicated, and $\epsilon=0.07$. 
Interestingly, in the region of unstable black holes (dashed lines),  $c_s^2$ becomes negative, meaning that the speed of sound becomes imaginary. 
This is related to the fact that the specific heat is always negative in this region. To be more specific we may rewrite the speed of sound in a convenient form, 
\noindent
\begin{equation} \label{Eq:cs2v2}
c_s^2=\frac{s}{C_V}.
\end{equation}
\noindent
Since the entropy density is always positive, c.f. Fig.~\ref{Fig1:zhTFAP}, a negative value of $C_V$ gives rise to a negative value for $c_s^2$, hence the results found for the small black holes on the left and right panel of Fig.~\ref{Fig1:TCVFAP} (dashed lines) are consistent. 
We then conclude that the instability of the small black hole is characterized by a negative specific heat and a purely imaginary speed of sound. 
Our numerical results also show that at $T=T_c$,  $c_s^2$ crosses the horizontal axis ($c_s^2=0$)  whereas  $C_V$ diverges. This is consistent with our formula \eqref{Eq:cs2v2}. 
Note that close to $T=T_{c}$, $c_s^2$ increases rapidly with $T$ and far from $T_c$ it varies slowly. 
Last but not least, we conclude that the speed of sound is also sensitive to the variation of $\hat\phi_0$. For the large (small) black hole solution  $c_s^2$ decreases (increases) for increasing $\hat\phi_0$. 
Note also that $c_s^2$ converges to $1/3$ for the (physical) large black holes in the limit of very high temperatures, which is consistent with the restoration of conformal symmetry.

\begin{figure}[ht!]
\centering
\includegraphics[width=7cm]{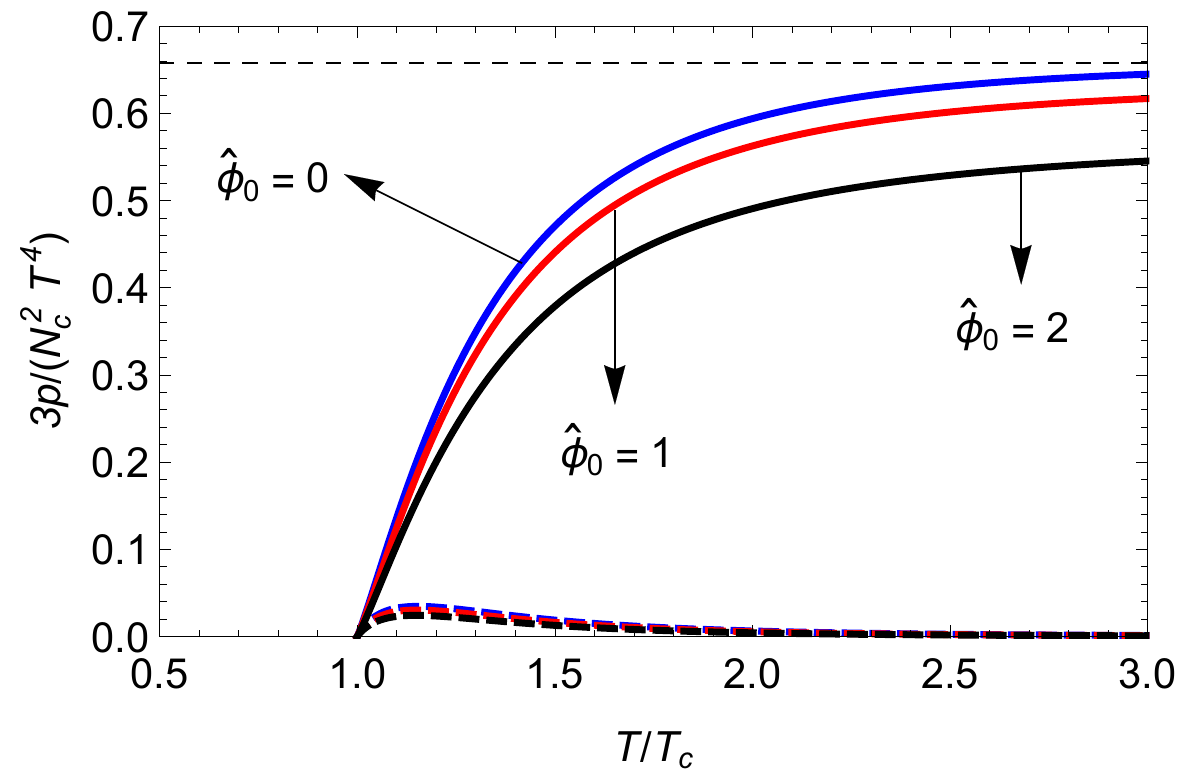}\\
\caption{
The dimensionless pressure $3p/(N_{c}^{2}T^4)$ as a function of the dimensionless temperature $T/T_c$ for three different values of $\hat\phi_0$, as indicated, and $\epsilon=0.07$. Solid (dashed) lines represent   the large (small) black holes regime. The thin dashed horizontal line represents the value $3p/(N_c^2 T^3)=3\pi^2/45$.
}
\label{Fig1:TFFAP}
\end{figure}

Using our formula \eqref{Eq:FreeEnergy} for the free energy density, we can evaluate the pressure $P = - F$. 
Our numerical results for the dimensionless pressure $3 p/(N_c^2T^4)$ as a function of the dimensionless temperature $T/T_c$ are displayed in Fig.~\ref{Fig1:TFFAP} for $\epsilon=0.07$ and different values of $\hat \phi_0$. 
Note that the sensitivity to the parameter $\hat\phi_0$ depends on the region of interest.  
In the branch describing small black holes (dashed lines), the pressure is less sensitive to $\hat{\phi}_0$ far from the critical temperature $T_c$. 
Meanwhile, in the branch of large black holes (solid lines), the pressure is more sensitive to $\hat{\phi}_0$ for temperatures much higher than $T_c$.

\begin{figure}[ht!]
\centering
\includegraphics[width=7cm]{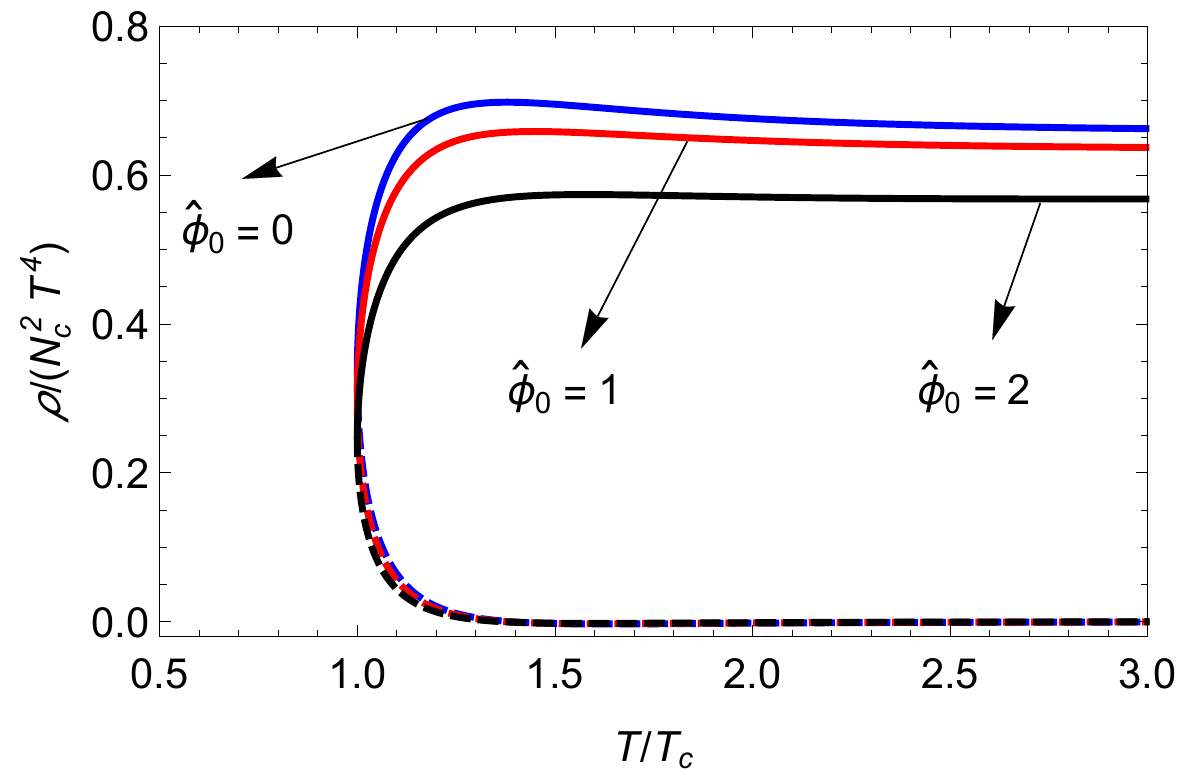}\hfill
\includegraphics[width=7cm]{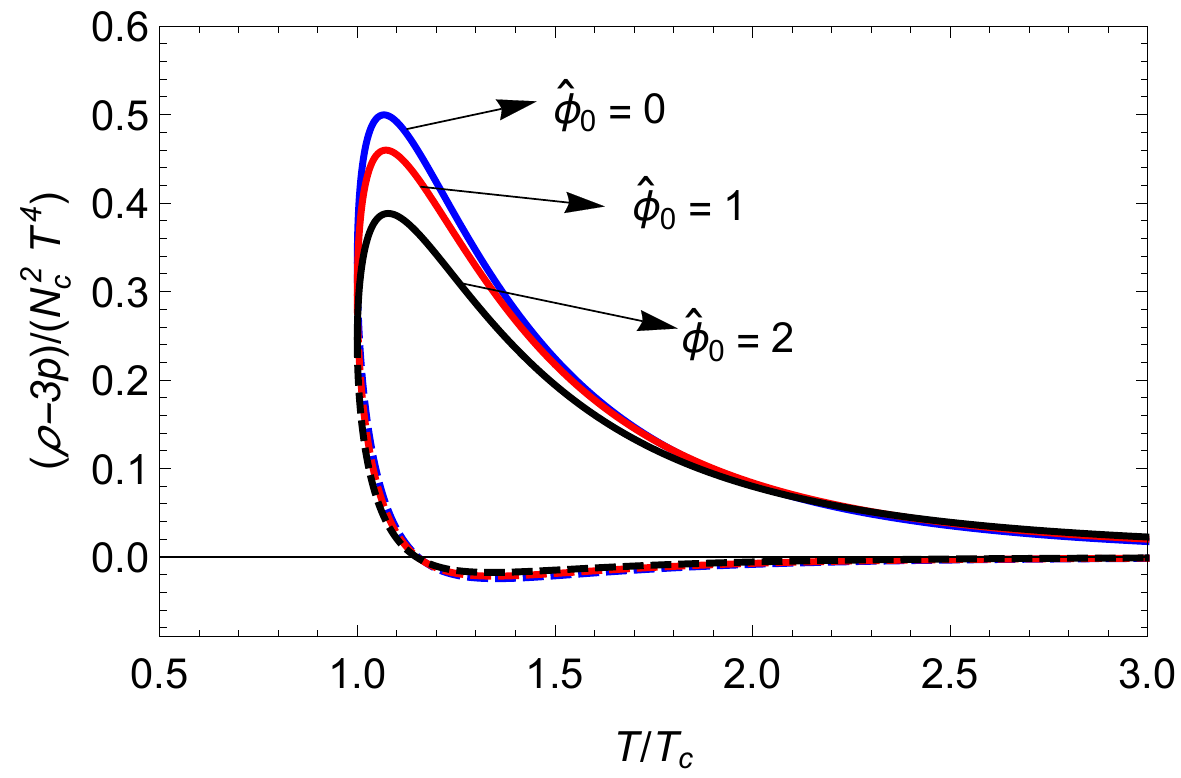}\\
\caption{
Left: The energy density $\rho/(N_c^2 T^4)$ as a function of $T/T_c$ for three different values of $\hat\phi_0$, as indicated, and $\epsilon=0,07$. Right: The trace anomaly $(\rho-3p)/(N_{c}^{2}T^4)$ (right panel)  as a function of $T/T_c$ for three different values of $\hat\phi_0$, as indicated, and $\epsilon=0,07$. Solid (dashed) lines represent   the large (small) black holes regime.
}
\label{Fig1:TEFAP}
\end{figure}

Our numerical results for the energy density $\rho$ as a function of the temperature are displayed on the left panel of Fig.~\ref{Fig1:TEFAP}, where we plot the quantity $\rho/(N_c^2T^4)$ for $\epsilon=0.07$ and different values of $\hat \phi_0$. 
As the figure shows, the energy density is also sensitive to $\hat\phi_0$, it decreases with the increasing of $\hat\phi_0$ for the large and small black hole regimes (solid and dashed lines respectively). We also observe that the energy density increases rapidly with $T$ close to the critical temperature $T_c$ and varies slowly in the regime of high temperatures. 
The right panel of Fig.~\ref{Fig1:TEFAP} displays our results for the dimensionless trace anomaly, $(\rho-3p)/(N_c^2T^4)$, as a function of the dimensionless temperature $T/T_c$ for $\epsilon=0.07$ and different values of $\hat\phi_0$. We can see that the trace anomaly decreases with the increasing of $\hat\phi_0$. 
Note that for the large black hole regime (solid lines) the trace anomaly has a peak close to the critical temperature, then, it decreases and goes to zero in the limit of very high temperatures recovering conformal symmetry. For the small black hole regime (dashed lines) the trace anomaly becomes negative as $T$ increases and goes to zero in the high temperature (conformal) limit. 

The main conclusion of the above analysis is as follows. When the conformal anomalous dimension $\epsilon$ is fixed, all the thermodynamic variables describing the non-conformal plasma in EHQCD are sensitive to the variation of the dimensionless coupling $\hat\phi_0$. 
We did a similar analysis for the case where we fix $\hat \phi_0$ and found that the thermodynamic variables are sensitive to the variation of $\epsilon$. 

Other important lessons that can be learned from the analysis presented in this section are the following. 
Holographic QCD models that describe CFT deformations consistent with confinement at zero temperature lead to the formation of non-conformal plasmas only above a certain temperature $T_c$. 
In other words, there is a minimum temperature for gluon deconfinement due to the CFT deformation. 
Moreover, above $T_c$ the black hole solution splits into two branches: a large stable black hole characterized by a positive specific heat and a real speed of sound and a small black hole characterized by a negative specific heat and an imaginary speed of sound. 
In our bottom-up framework, CFT deformations have been described by an effective 5d Einstein-dilaton theory. 
Interestingly, qualitatively similar results can be obtained considering a top-down approach, e.g. \cite{Mamo:2016dew, Mamo:2016oli}

\section{Viscosity coefficients of the non-conformal plasma}
\label{Sec:ViscosityCoeffs}

There are at least three methods for calculating transport coefficients of a fluid arising from a strongly coupled theory within the framework of holography. 
One of them is through the hydrodynamic limit of the black hole quasinormal modes, where one calculates the dispersion relations of field perturbations on the gravitational side of the duality \cite{Policastro:2002se, Kovtun:2005ev}. 
These relations are then compared to the ones obtained from the linearized hydrodynamic modes. 
In general, the dispersion relations arising from the relativistic hydrodynamics are constructed by using the gradient expansion method (see, for instance, Refs.~\cite{Baier:2007ix, Kovtun:2012rj, Grozdanov:2015kqa, Diles:2019uft}). 
Notice that in order to calculate the viscosity coefficients we need to consider first-order hydrodynamics.

A second method consists of the direct evaluation of the retarded Green's functions for the components of the stress-energy tensor associated with the transport coefficients. 
Using the AdS/CFT correspondence, one identifies the dual 5d gravitational perturbations. 
Then, the action is expanded up to second-order in the perturbations to read off the retarded Green's function. 
Finally, Kubo's formula is used to calculate the desired transport coefficients. 
This is the approach used in the seminal work \cite{Kovtun:2004de} for the shear viscosity and \cite{Gubser:2008sz} for the bulk viscosity. 

A third method is based on the fluid/gravity correspondence \cite{Hubeny:2011hd}, where the full stress-energy tensor can be obtained from the 5d metric by solving the full non-linear Einstein equations via a gradient expansion. 
In this paper, we will follow the second method described above to calculate the shear and bulk viscosities of the non-conformal plasma arising in our EHQCD model.

\subsection{Shear viscosity}
\label{SubSec:Shear}

In the case of the shear viscosity ($\eta$), considering the direction of propagation $k^{\mu}=(\omega,0,0,q)$, the relevant retarded Green's function is given by
\noindent
\begin{equation}
G_R(\omega,q)=-i\int dt d^3x\,
e^{i\omega t-iq\,x^3}
\theta(t)\langle[T_{x^1x^2}(t,x^{3}),T_{x^1x^2}(0,0)]\rangle,
\label{GreenFunc}
\end{equation}
\noindent
where $T_{x^1x^2}$ is one of the energy-momentum tensor components and $\omega$ ($q$) is the frequency (wavenumber) of the perturbation. 
From the holographic dictionary, we know that the source of the shear viscosity is related to the metric perturbation $h_{x^{1}x^{2}}$, which couples to the component $T_{x^1x^2}$ of the energy-momentum tensor, relevant for calculating $\eta$.

In the following, we develop a general procedure to calculate the shear viscosity in the holographic model we are working with. 
Let us start with a general ansatz for the black hole metric in holographic QCD, which we write as
\noindent
\begin{equation}\label{Eq:MetricTwoWarpFactors}
ds^2=\frac{1}{\zeta_1^2}\left(-f\,dt^2+dx_idx^{i}\right)+\frac{dz^2}{\zeta_2^2f},
\end{equation}
\noindent
where $f$, $\zeta_1$, and $\zeta_2$ are functions of $z$ only. We use this ansatz for the metric in order to compare with previous 
results in the literature and check the consistency of our procedure and results. The corresponding background equations are
\noindent
\begin{equation}\label{EqsBH2}
\begin{split}
\frac{\zeta'_1\,\zeta'_2}{\zeta_1\,\zeta_2}-\frac{\zeta'^{\,2}_{1}}{\zeta_1^2}+\frac{\zeta''_1}{\zeta_1}
-\frac{4}{9}\Phi'^{\,2}=&\,0,\\
3\zeta''_1
+3\zeta'_1\left(\frac{f'}{f}-5\frac{\zeta'_1}{\zeta_1}+\frac{\zeta'_2}{\zeta_2}\right)+\frac{\zeta_1\,V}{f\,\zeta_2^2}=&\,0,\\
\frac{f''}{f'}-4\frac{\zeta'_1}{\zeta_1}+\frac{\zeta'_2}{\zeta_2}=&\,0.
\end{split}
\end{equation}
\noindent
As expected, these relations reduce to \eqref{EqsBH} for $\zeta_1=\zeta_2$. 

In order to calculate the shear viscosity, the next step is to consider a perturbation on the background black hole metric, $g_{mn}\to g_{mn}+h_{mn}$. 
To calculate the shear viscosity we need to consider the component $h_{x^{1}x^{2}}$ only, and this perturbation decouples naturally from the others. 
The relevant equation governing this sector is obtained from
\noindent
\begin{equation}
G_{x^{1}x^{2}}^{\scriptscriptstyle{(1)}}=\frac{1}{2\,\sigma}T_{x^{1}x^{2}}^{\scriptscriptstyle{(1)}},
\end{equation}
\noindent
where $G_{x^{1}x^{2}}^{\scriptscriptstyle{(1)}}$ and $T_{x^{1}x^{2}}^{\scriptscriptstyle{(1)}}$ are first-order contributions to the Einstein and energy-momentum tensor expansions in $h_{x^{1}x^{2}}$. 
Since $x^3$ is the direction of propagation of fluctuations in the transverse space, the differential equation for $h_{x^1x^2}(t,x^3,z)$ may be written as
\noindent
\begin{equation}
\begin{split}
\partial^2_z h_{x^{1}x^{2}}+\left(\frac{f'}{f}+\frac{\zeta'_2}{\zeta_2}\right)\partial_z h_{x^{1}x^{2}}&+\left(\frac{2f'\,\zeta'_1}{f\,\zeta_1}-\frac{4\zeta'^{\,2}_{1}}{\zeta_1^2}+
\frac{8}{9}\Phi'^{\,2}\right)h_{x^{1}x^{2}}\\ &+\frac{\zeta_1^2}{f^2\zeta_2^2}\left(f\,\partial^2_{x^3} h_{x^{1}x^{2}}-
\partial^2_{t} h_{x^{1}x^{2}}\right)=0.
\end{split}
\end{equation}
\noindent
However, this equation is not gauge-invariant. As discussed in Refs. \cite{Policastro:2002se, Kovtun:2005ev} we may write it in terms of a gauge-invariant master field, 
the so-called  Kovtun-Starinets (KS) master variable, defined by $Z_T=g^{x^{1}x^{1}}h_{x^{1}x^{2}}$. In doing so, we obtain
\noindent
\begin{equation}
\partial^2_z Z_{T}+\left(\frac{f'}{f}-\frac{f''}{f'}\right)\partial_z Z_{T}+\frac{\zeta_1^2}{f^2\zeta_2^2}
\left(f\,\partial^2_{x^3} Z_{T}-\partial^2_{t} Z_{T}\right)=0.
\end{equation}
\noindent
Considering the Fourier transform
\noindent
\begin{equation}
Z_T(t,x^3,z)=\int \frac{dq d\omega}{(2\pi)^2}e^{-i \omega t+i q x^{3}}Z_T(\omega,q,z),
\end{equation}
\noindent
the fundamental equation becomes
\noindent\begin{equation}\label{Eq:KSMasterEq}
Z_{T}''(z)-\left(\ln{\left[\frac{f'}{f}\right]}\right)'Z_{T}'(z)+
\frac{\zeta_1^2}{f^2\zeta_2^2}\left(\omega^2-q^2f\right)Z_{T}(z)=0.
\end{equation}
\noindent
At this point, it is interesting to observe that considering $\zeta_1$ and $\zeta_2$ appropriately, equation \eqref{Eq:KSMasterEq} reduces to previous results presented in the literature. 
For example, in the conformal case, the dilaton field is zero and one may compare this equation with Eq.~(6.6) of Ref.~\cite{Policastro:2002se}. 
It is also interesting mentioning that equation \eqref{Eq:KSMasterEq} takes the same form as Eq.~(2.11) of Ref.~\cite{Gubser:2008sz}, setting $q=0$. 
Finally, following the procedure implemented in Ref.~\cite{Morgan:2009pn}, we may write this equation in terms of the Regge-Wheeler-Zerilli master variable \footnote{We shall address this approach in the future.}.

To find the retarded Green's function \eqref{GreenFunc} we now expand the action up to second-order in the fluctuation, $h_{x^1x^2}$. 
Then, we write the resulting on-shell action in terms of the gauge-invariant field $Z_T$ in the form
\noindent
\begin{equation}
S^{\scriptscriptstyle{(2)}}=\sigma\int d^5x\mathcal{L}^{\scriptscriptstyle{(2)}},
\end{equation}
where the Lagrangian is given by
\noindent
\begin{equation}
\mathcal{L}^{\scriptscriptstyle{(2)}}=\mathcal{\widehat{L}}^{\scriptscriptstyle{(2)}}+
\partial_{t}\mathcal{L}^{t}+\partial_{x^3}\mathcal{L}^{x^3}+\partial_{z}\mathcal{L}^{z},
\end{equation}
\noindent
with
\noindent
\begin{equation}
\begin{split}
\mathcal{\widehat{L}}^{\scriptscriptstyle{(2)}}=&\frac{1}{2f\,\zeta_1^2\zeta_2}(\partial_{t}Z_T)^2-
\frac{1}{2\zeta_1^2\zeta_2}(\partial_{x^3}Z_T)^2-\frac{f\zeta_2}{2\zeta_1^4}Z_T'^{\,2},\\
\mathcal{L}^{{t}}=&\frac{-2}{f\zeta_1^2\zeta_2}Z_T\,\partial_{t}Z_T,\\
\mathcal{L}^{{x^3}}=&\frac{2}{\zeta_1^2\zeta_2}Z_T\,\partial_{x^3}Z_T,\\
\mathcal{L}^{{z}}=&\frac{2f\zeta_2}{\zeta_1^4}Z_T\,Z_T'-\frac{f\zeta_2\zeta_1'}{\zeta_1^5}Z_T^2.
\end{split}
\end{equation}
\noindent

As discussed in Ref.~\cite{Gubser:2008sz}, it is possible to include the contribution of the Gibbons-Hawking-York surface term to the Lagrangian $\mathcal{\widehat{L}}^{\scriptscriptstyle{(2)}}$ by adding a non-trivial scalar function $G(z)$, such that we rule out contributions of the form $Z_T^2$ and $Z_T\, Z_T'$. 
In turn, as $G(z)$ is an arbitrary (unknown) function, this will generate an ambiguity. However, the profile of $G(z)$ may be fixed by phenomenology, eliminating the aforementioned ambiguity. 
Then, adding the contribution of this function to the Lagrangian we get
\noindent
\begin{equation}
\begin{split}
\mathcal{\widehat{L}}^{\scriptscriptstyle{(2)}}=\frac{1}{2f\,\zeta_1^2\zeta_2}\left(\partial_{t}Z_T\right)^2-
\frac{1}{2\zeta_1^2\zeta_2}\left(\partial_{x^3}Z_T\right)^2-\frac{f\zeta_2}{2\zeta_1^4}Z_T'^{\,2}+\frac{1}{2}\partial_z\left(G\, Z_T^2\right).
\end{split}
\end{equation}
\noindent
Nevertheless, the additional term of the ``improved'' Lagrangian will not contribute to the imaginary part of Green's function. 
As we shall see, the shear viscosity coefficient depends on the imaginary part of Green's function; 
then, the explicit form of $G(z)$ does not matter for our calculation.

The next step forward is to promote $Z_T$ to a complex function, and this is because, in general, an arbitrary solution of Eq.~\eqref{Eq:KSMasterEq} might be complex, depending on the boundary conditions of the problem. Thus, we rewrite the Lagrangian using the master variable and its complex conjugate,
\noindent
\begin{equation}\label{Eq:ImprLagrShear}
\begin{split}
\mathcal{\widehat{L}}_{\bf{C}}^{\scriptscriptstyle{(2)}}\equiv\frac{\omega^2}{f\,\zeta_1^2\zeta_2}|Z_T|^2-
\frac{q^2}{\zeta_1^2\zeta_2}|Z_T|^2-\frac{f\zeta_2}{\zeta_1^4}|Z_T'|^2+G' |Z_T|^2+G\left(Z_T {Z_T^{*}}'+Z_T^{*} Z_T'\right).
\end{split}
\end{equation}
\noindent
The form of this Lagrangian provides the same equation of motion \eqref{Eq:KSMasterEq}. It is also possible to rewrite the last Lagrangian in a form where the presence of a surface term is evident,
\begin{equation}\label{Eq:ImprLagrrShearb}
\mathcal{\widehat{L}}_{\bf{C}}^{\scriptscriptstyle{(2)}}=\partial_{z}J+Z_T^{*}\left(\frac{\partial \mathcal{\widehat{L}}_{\bf{C}}^{\scriptscriptstyle{(2)}}}{\partial Z_T^{*}}-\frac{d}{dz}\frac{\partial \mathcal{\widehat{L}}_{\bf{C}}^{\scriptscriptstyle{(2)}}}{\partial {Z_T^{*}}'}\right),
\end{equation}
\noindent
where 
\begin{equation}
J=-\frac{f\zeta_2}{\zeta_1^4}Z_T^{*}Z_T'+G Z_T^{*}Z_T.
\end{equation}
\noindent
At the boundary, $J$ must be related to the retarded Green's function. To calculate the imaginary part of the Green's function \eqref{GreenFunc} we need to define the number flux of gravitons, $\mathcal{F}$, which is related to the imaginary part of $J$ through
\noindent
\begin{equation}\label{Eq:FluxNumberA}
\mathcal{F}=-{\rm Im}\,J=\frac{1}{2i}\frac{f\zeta_2}{\zeta_1^4}\left(Z_T^{*}Z_T'-Z_T {Z_T^{*}}'\right).
\end{equation}
\noindent
As described in \cite{Gubser:2008sz}, the quantity $\mathcal{F}$ represents the number flux of gravitons in the radial direction and it is the conserved charge associated with the $U(1)$ symmetry of \eqref{Eq:ImprLagrShear}. We shall use this expression in the last part of this section. 

Let us now focus on the case we are dealing with, where $\zeta_1=\zeta_2$. To calculate the shear viscosity we follow the procedure implemented in Ref.~\cite{Gubser:2008sz}. First, we solve Eq.~\eqref{Eq:KSMasterEq} in the limit of zero frequency and wavenumber, i.e. $(\omega,q)\to 0$, where the resulting differential equation has an exact solution, given by
\noindent
\begin{equation}\label{Eq:ZT1}
Z_T= C_2+C_1\ln{f}.
\end{equation}
\noindent
Here $C_1$ and $C_2$ are integration constants, and in the limit $\omega \to 0$ we must set $C_1=0$ in order to guarantee the regularity condition at the horizon, where $f(z_h)=0$. Without loss of generality, we also impose the Dirichlet boundary condition $Z_T=1$ at the boundary, so that $C_2=1$. 
Now consider the case of small $\omega$ (with $q=0$) where we  solve Eq.~\eqref{Eq:KSMasterEq} perturbatively, considering $\omega$ as the expansion parameter. In that case, the solution takes the same form as Eq.~\eqref{Eq:ZT1}, where $C_1(\omega)$ depends on the frequency. 
Then, expanding this solution close to the horizon we obtain the approximate solution
\noindent
\begin{equation}\label{Eq:ZTHorizon}
Z_T\approx C_2+C_1\ln{(z_h-z)}.
\end{equation}
\noindent

\begin{figure}[ht!]
\centering
\includegraphics[width=7cm]{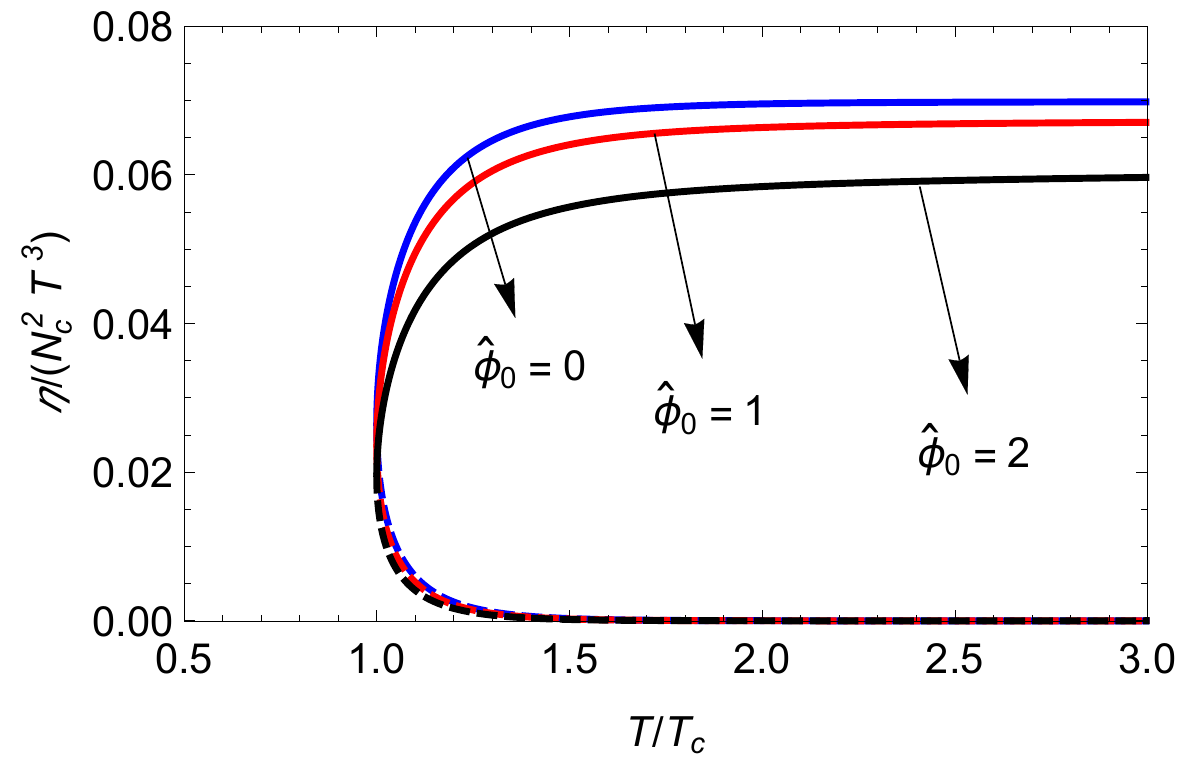}
\caption{
The normalized shear coefficient $\eta/(N_c^2\,T^3)$ as a function of $T/T_c$ for different values of $\hat\phi_0$ and $\epsilon=0.07$. Solid (dashed) lines indicate the large (small) black hole regime.
}
\label{Fig1:T3EtaFAP}
\end{figure}

The next stage is to solve the differential equation \eqref{Eq:KSMasterEq} close to the horizon, for arbitrary $\omega$ and $q$. To do so, we use the ansatz $Z_T=f^{\beta}$, where $\beta$ takes the values: 
\noindent
\begin{equation}
\beta_1=\frac{i\omega}{f'(z_h)}, \qquad\qquad \beta_2=-\frac{i\omega}{f'(z_h)},
\end{equation}
\noindent
where $f'(z_h)=-4 \pi T$ with $T$ being the black hole temperature. We choose the solution associated with $\beta_1$ because it represents waves falling into the black hole. 
This condition is also related to the retarded Green's function. Then, considering the first subleading term close to the horizon, $Z_T$ is given by
\noindent
\begin{equation}
Z_T=F_0f^{-\frac{i\omega}{4\pi T}}\left[1+\frac{i(q^2 f'^2(z_h)+2\omega^2
f''(z_h))}{(2\omega-if'(z_h))f'^2(z_h)}(z_h-z)+\cdots\right],
\end{equation}
\noindent
or, expanding $f(z)$ around $z=z_h$,
\noindent
\begin{equation}
Z_T= \mathcal{A}_{-} (z_h-z)^{-\frac{i\omega}{4\pi T}}\left[1+\frac{i(q^2f'^2(z_h)+2\omega^2f''(z_h))}{(2\omega-if'(z_h))f'^2(z_h)}(z_h-z)+\cdots\right],
\label{Eq:SolHorizon}
\end{equation}
\noindent
where $F_0$ and $\mathcal{A}_{-}$ are constants. Considering $q=0$ and expanding the leading term around $\omega=0$, we get
\noindent
\begin{equation}\label{Eq:ZT2}
Z_T\approx \mathcal{A}_{-}\left[1-\frac{i\omega}{4\pi T}\ln{(z_h-z)}\right].
\end{equation}
\noindent
This approximate solution must be equal to Eq.~\eqref{Eq:ZTHorizon}. Thus, we identify the corresponding coefficients:
\noindent
\begin{equation}
\mathcal{A}_{-}=C_2=1,\qquad C_1=-\frac{i \omega \mathcal{A}_{-}}{4\pi T}.
\end{equation}
\noindent
Note that $C_1$ vanishes in the limit $\omega \to 0$, as expected. Having found the asymptotic solution for $Z_T$ near the horizon in \eqref{Eq:SolHorizon} we can evaluate the (conserved) graviton number flux $\mathcal{F}$ in the limit $z \to z_h$.  
Plugging the result \eqref{Eq:SolHorizon} for $Z_T$ into the  number flux formula \eqref{Eq:FluxNumberA} we get (setting $q=0$)
\noindent
\begin{equation}
\mathcal{F}=\frac{\omega}{\zeta_1^3(z_h)}.
\end{equation}
\noindent
The imaginary part of the retarded Green's function takes the form 
\noindent
\begin{equation}
{\rm{Im}}\, G_R(\omega,q= 0)=-\frac{\mathcal{F}}{16\pi G_5}=-\frac{\omega}{16\pi G_5\zeta_1^3(z_h)}.
\end{equation}
\noindent
Finally, we extract the shear viscosity using the Kubo's formula:
\noindent
\begin{equation}\label{Eq:ShearVescosity}
\eta=-\lim_{\omega\to 0}\frac{1}{\omega} {\rm{Im}}\, G_R(\omega,q=0)=\frac{1}{16\pi G_5\zeta_1^3(z_h)}.
\end{equation}
\noindent
Just like the entropy density, the shear viscosity coefficient $\eta$ depends on the (inverse) scale factor evaluated at the horizon. In Fig.~\ref{Fig1:T3EtaFAP} we display our numerical results for the shear viscosity, normalized as $\eta/(N_c^2\,T^3)$, as a function of the dimensionless temperature $T/T_c$ for different values of the dimensionless coupling $\hat\phi_0$, for a fixed value $\epsilon=0.07$ of the conformal anomalous dimension. 
We observe that close to the critical temperature the shear viscosity increases rapidly with the temperature whereas far from the critical temperature it varies slowly. We conclude that the shear viscosity is also sensitive to the value of $\hat\phi_0$, for fixed $\epsilon$.

It is worth pointing out that the ratio $\eta/s$ does not change under the variation of $\hat \phi_0$ or $\epsilon$. This can be shown by plugging \eqref{Eq:Entropy} in \eqref{Eq:ShearVescosity}, thus, we get the famous result
\noindent
\begin{equation}
\frac{\eta}{s}=\frac{1}{4\pi}.
\end{equation}
\noindent
This ratio is expected to change when higher curvature terms are added in the five-dimensional action, see for instance \cite{Cremonini:2012ny}. We would like to remark that, in contrast with  Ref.~\cite{Gubser:2008sz}, we did not need to redefine our radial coordinate $z$ in terms of the dilaton field $\Phi$. 
This is true whenever we express the perturbation in terms of a gauge-invariant master field. In conclusion, the master field $Z_T(z)$ is equivalent to the field $H_{12}(\Phi)$ used in Ref.~\cite{Gubser:2008sz}.

\subsection{Bulk viscosity}
\label{SubSec:Bulk}

Following the conventions of Ref.~\cite{Gubser:2008sz}, the retarded Green's function related to the bulk viscosity is given by
\noindent
\begin{equation}
G_{R}(\omega,q)=-i\int dt\, d^3 x \,e^{i\omega\,t-iqx^{3}}\,\theta(t)\left\langle\left[\frac12T^{i}_{i}(t,x^3),\frac12 T^{k}_{k}(0,0)\right]\right\rangle,
\label{Green_function}
\end{equation}
\noindent
where $T_{i}^{i}$ is the trace of the spatial part of the energy-momentum tensor. The holographic dictionary then indicates that the source of the bulk viscosity is related to the trace of the metric perturbations.

The bulk viscosity has been investigated previously in holography (see, for instance, Refs.~\cite{Buchel:2005cv, Buchel:2007mf,Gubser:2008sz, Springer:2008js, Gursoy:2009kk, Buchel:2011uj, Eling:2011ms, Buchel:2011wx, Finazzo:2014cna, Yaresko:2013tia, Czajka:2018egm, Li:2014dsa}). This transport coefficient is associated with the expansion scalar term $\partial_\mu u^{\mu}$ of the fluid stress-energy tensor and associated with conformal symmetry breaking. 
Here we follow the procedure implemented in Ref.~\cite{Gubser:2008sz} adapted to our case. The gauge where the dilaton field becomes the holographic coordinate simplifies dramatically the problem of calculating this transport coefficient. 
Such a choice is possible whenever the dilaton is a monotonic increasing function. Thus, within this gauge the metric \eqref{Eq:MetricTwoWarpFactors} becomes
\noindent
\begin{equation}\label{Eq:MetricRadialDil}
ds^2=\frac{1}{\zeta_1^2}\left(-f\,dt^2+dx_idx^{i}\right)+\frac{d\Phi^2}{\zeta_2^2f}.
\end{equation}
\noindent
It is worth pointing out that the metric \eqref{Eq:MetricRadialDil} reduces to \eqref{EqBHmetric} considering $\zeta_2=\Phi'(z)\zeta_1$. In this gauge, we do not need to care about the perturbation of the scalar field, which naturally shall be coupled to the trace of the metric perturbations. 
We restrict ourselves to the zero spatial momentum case ($q=0$); thus, the perturbed metric depends on the time and dilaton field only, and the metric components are given in explicit form by
\begin{equation}
g_{mn}={\rm{diag}}\left(g_{tt},\, g_{x^1x^1},\,g_{x^1x^1},\,g_{x^1x^1},\,g_{\Phi\Phi}\right),
\end{equation}
where 
\noindent
\begin{equation}\label{Eq:MetricBulk}
\begin{split}
g_{tt}=&-\frac{f}{\zeta_1^2}\left(1+\frac{\lambda}{2} H_{00}(t,\Phi)\right),\\
g_{x^1x^1}=&\frac{1}{\zeta_1^2}\left(1+\frac{\lambda}{2} H_{11}(t,\Phi)\right),\\
g_{\Phi\Phi}=&\frac{1}{\zeta_2^2f}\left(1+\frac{\lambda}{2} H_{55}(t,\Phi)\right).
\end{split}
\end{equation}
\noindent
Here $\zeta_2$, $\zeta_1$ and $f$ are functions of the dilaton field, whereas $\lambda$ ($\lambda\ll 1$)
is a small parameter introduced to control the expansion. In turn, the background equations are:
\noindent
\begin{equation}\label{EqsBH3}
\begin{split}
\frac{\partial_{\Phi}\zeta_1}{\zeta_1}\frac{\partial_{\Phi}\zeta_2}{\zeta_2}-\frac{(\partial_{\Phi}\zeta_1)^2}{\zeta_1^2}
+\frac{\partial_{\Phi}^2\zeta_1}{\zeta_1}-\frac{4}{9}=&\,0,\\
f\left(12\frac{\left(\partial_{\Phi}\zeta_1\right)^2}{\zeta_1^2}-\frac{4}{3}\right)
-3\frac{(\partial_{\Phi}f)\,(\partial_{\Phi}\zeta_1)}{\zeta_1}-\frac{V}{\zeta_2^2}=&\,0,\\
\frac{\partial_{\Phi}^{2}f}{\partial_{\Phi}f}-4\frac{\partial_{\Phi}\zeta_1}{\zeta_1}+\frac{\partial_{\Phi}\zeta_2}{\zeta_2}=&\,0.
\end{split}
\end{equation}
\noindent

Plugging \eqref{Eq:MetricBulk} into the Einstein equations \eqref{EqEinstein}, and then performing a Fourier transform, we get the corresponding perturbation equations:
\noindent
\begin{subequations}
\begin{align}
0=&-\frac{\partial_{\Phi}\ln{f}}{2\,\partial_{\Phi}\ln{\zeta_1}}H_{11}-\frac{1}{\partial_{\Phi}\ln{\zeta_1}}
\partial_{\Phi}H_{11}+H_{55}, \label{Eq:1}\\%
\begin{split}
0=&\left(\frac{(\partial_{\Phi}\ln{f})^2}{2\,\partial_{\Phi}\ln{\zeta_1}}-2\,\partial_{\Phi}\ln{f}+\frac{2\,
\partial_{\Phi}\ln{f}}{9\,(\partial_{\Phi}\ln{\zeta_1})^2}+\frac{\omega^2\zeta_1^2}{f^2\zeta_2^2(\partial_{\Phi}
\ln{\zeta_1})}\right)H_{11}\\
&+\left(1-\frac{4}{9(\partial_{\Phi}\ln{\zeta_1})^2}-\frac{\partial_{\Phi}\ln{f}}{2\,
\partial_{\Phi}\ln{\zeta_1}}\right)\partial_{\Phi}H_{11}-\partial_{\Phi}H_{00}, \label{Eq:2}
\end{split}\\
\begin{split}
0=&\left((\partial_{\Phi}\ln{f})(\partial_{\Phi}\ln{\zeta_2)}-\frac{4\,\partial_{\Phi}
\ln{f}}{9\,\partial_{\Phi}\ln{\zeta_1}}-\frac{\omega^2\zeta_1^2}{f^2\zeta_2^2}\right)H_{11}\\
&+\left(\frac{8}{9\,\partial_{\Phi}\ln{\zeta_1}}+\partial_{\Phi}
\ln{\left[\frac{\zeta_1^4}{\zeta_2^3f}\right]}\right)\partial_{\Phi}H_{11}-\partial_{\Phi}^2H_{11}.\label{Eq:3}
\end{split}
\end{align}
\end{subequations}
\noindent
As can be seen from the equations, once we solve the differential equation \eqref{Eq:3} we may automatically calculate $H_{00}$ and $H_{55}$. Using the background differential equations \eqref{EqsBH3} we may rewrite Eq.\eqref{Eq:3} in the reduced form
\noindent
\begin{equation}\label{Eq:3Improved}
\partial_{\Phi}^2H_{11}-\partial_{\Phi}\ln{\left[(\partial_{\Phi}\ln{f})\left(\partial_{\Phi}\ln{\zeta_1}\right)^2\right]}
\partial_{\Phi}H_{11}+\left[\frac{\omega^2\zeta_1^2}{f^2\zeta_2^2}+\left(\partial_{\Phi}\ln{f}\right)\left(\partial_{\Phi}
\ln{[\partial_{\Phi}\ln{\zeta_1}]}\right)\right]H_{11}=0.
\end{equation}

To obtain the bulk viscosity we need to calculate the retarded Green's function related to these metric perturbations. Thus, analogously to what has been done with the shear viscosity, 
we substitute the metric \eqref{Eq:MetricBulk} in the action  \eqref{EqAction} and then expand up to second-order in $\lambda$. The resulting on-shell action may be written as
\noindent
\begin{equation}\label{Eq:H11}
S^{\scriptscriptstyle{(2)}}=\sigma\int d^5x\,\mathcal{L}^{\scriptscriptstyle{(2)}}.
\end{equation}
The Lagrangian is given by
\noindent
\begin{equation}
\mathcal{L}^{\scriptscriptstyle{(2)}}=\mathcal{\widehat{L}}^{\scriptscriptstyle{(2)}}+
\partial_{t}\mathcal{L}^{t}+\partial_{\Phi}\mathcal{L}^{\Phi},
\end{equation}
\noindent
where
\begin{equation}
\begin{split}
\mathcal{\widehat{L}}^{\scriptscriptstyle{(2)}}=&\frac{1}{2}\partial_t\vec{H}^{T}{\bf M}^{tt}\partial_t\vec{H}+\frac{1}{2}\partial_\Phi\vec{H}^{T}{\bf M}^{\Phi\Phi}\partial_\Phi\vec{H}+\frac{1}{2}\vec{H}^{T}{\bf M}
\vec{H}+\frac{1}{2}\partial_{\Phi}\vec{H}^{T}{\bf M}^{\Phi}\vec{H}.
\end{split}
\end{equation}
\noindent
To get this Lagrangian we have used the equality
$\partial_{\Phi}\vec{H}^{T}{\bf M}^{\Phi}\vec{H}=\vec{H}^{T}({\bf M}^{\Phi})^{T}\partial_{\Phi}\vec{H}$.
The matrices are given by
\noindent
\begin{equation}\label{Eq:Matrices}
\begin{split}
\vec{H}=&
\begin{pmatrix}
H_{00}\\
H_{11}\\
H_{55}
\end{pmatrix}, \quad 
{\bf M}^{tt}=-\frac{3}{\zeta_{1}^{2}\zeta_2 f}
\begin{pmatrix}
0 & 0 & 0\\
0 & 1 & \frac{1}{2}\\
0 & \frac{1}{2} & 0
\end{pmatrix},\quad 
{\bf M}^{\Phi\Phi}=\frac{3 f\zeta_2}{\zeta_1^4}
\begin{pmatrix}
0 & \frac12 & 0\\
\frac12 & 1 & 0\\
0 & 0 & 0
\end{pmatrix},\\
{\bf M}=&\frac{f\zeta_2}{3\zeta_1^4}\left[9\left(\partial_{\Phi}\ln{\zeta_1}\right)\left(\partial_{\Phi}\ln{f}\right)+
4\left(1-9\left(\partial_{\Phi}\ln{\zeta_1}\right)^2\right)\right]
\begin{pmatrix}
0 & \frac12 & 0\\
\frac12 & 1 & 0\\
0 & 0 & -\frac12
\end{pmatrix},\\
{\bf M}^{\Phi}=&\frac{3f\zeta_2(\partial_{\Phi}\ln{\zeta_1})}{\zeta_{1}^{4}}
\begin{pmatrix}
0 & -\frac{1}{2}\left(\frac{\partial_{\Phi}\ln{f}}{2(\partial_{\Phi}\ln{\zeta_1})}-1\right) & \frac{1}{2}\\
\frac{1}{2} & 1 & -\frac{1}{2}\left(\frac{\partial_{\Phi}\ln{f}}{2\partial_{\Phi}\left(\ln{\zeta_1}\right)}-3\right)\\
0 & 0 & 0
\end{pmatrix},
\end{split}
\end{equation}
whereas $\mathcal{L}^{\Phi}$ and $\mathcal{L}^{t}$ can be written as
\noindent
\begin{equation*}
\begin{split}
\mathcal{L}^{\Phi}=&-\frac{3f\zeta_2}{2\zeta_1^4}\partial_{\Phi}\left(H_{00}H_{11}\right)+\frac{f\zeta_2}{2\zeta_1^4}H_{55}\,
\partial_{\Phi}H_{00}-\frac{3f\zeta_2}{\zeta_1^4}H_{11}\,\partial_{\Phi}H_{11}+\frac{3f\zeta_2}{2\zeta_1^4}H_{55}\,\partial_{\Phi}H_{11}\\
&+\frac{\zeta_2f}{4\zeta_1^4}\left(\partial_{\Phi}\ln{f}-8\,\partial_{\Phi}\ln{\zeta_1}\right)H_{55}H_{00}+
\frac{3\zeta_2f}{4\zeta_1^4}\left(\partial_{\Phi}\ln{f}-8\,\partial_{\Phi}\ln{\zeta_1}\right)H_{55}H_{11}\\
&-\frac{\zeta_2f}{4\zeta_1^4}\left(\partial_{\Phi}\ln{f}-8\,\partial_{\Phi}\ln{\zeta_1}\right)H_{55}H_{55},\\
\mathcal{L}^{t}=&\frac{3}{f\zeta_1^2\zeta_2}H_{11}\,\partial_{t}H_{11}+\frac{3}{2f\zeta_1^2\zeta_2}\partial_{t}(H_{11}H_{55})-
\frac{3}{2f\zeta_1^2\zeta_2}H_{00}\,\partial_{t}H_{11}
-\frac{1}{2f\zeta_1^2\zeta_2}H_{00}\,\partial_{t}H_{55}.
\end{split}
\end{equation*}

In this case, ${\bf G}$ is a symmetric matrix whose elements may depend on the dilaton. Such a function has the same role as the one introduced in Eq.~\eqref{Eq:ImprLagrShear}. 
Thus, it will not contribute to the imaginary part of the retarded Green's function \eqref{Green_function}. In analogy to what was done in the previous section, we write down the improved Lagrangian
\noindent
\begin{equation}\label{Eq:ImprLagrrBulk}
\mathcal{\widehat{L}}_{{\bf C}}^{\scriptscriptstyle{(2)}}=\partial_{\Phi} \vec{H}^{*T}{\bf M}^{\Phi\Phi}\,\partial_{\Phi}\vec{H}+\vec{H}^{*T}{\bf K}\,\vec{H}+\partial_{\Phi} \vec{H}^{*T}{\bf B}\,\vec{H}+\vec{H}^{*T}{\bf B}^{*T}\,\partial_{\Phi}\vec{H},
\end{equation}
\noindent
where we have promoted the functions to complex ones and also defined
\begin{equation}
{\bf K}=\omega^2{\bf M}^{tt}+{\bf M}+\partial_{\Phi}{\bf G},\qquad {\bf B}={\bf M}^{\Phi}+{\bf G}.
\end{equation}
The next step is to rewrite this Lagrangian in an analogous form to Eq.~\eqref{Eq:ImprLagrrShearb}, where a surface term is evidenced. The result is
\begin{equation}\label{Eq:ImprLagrrBulkb}
\mathcal{\widehat{L}}_{\bf{C}}^{\scriptscriptstyle{(2)}}=\partial_{\Phi}J+\vec{H}^{*T}\left(\frac{\partial \mathcal{\widehat{L}}_{\bf{C}}^{\scriptscriptstyle{(2)}}}{\partial \vec{H}^{*T}}-\frac{d}{d\Phi}\frac{\partial \mathcal{\widehat{L}}_{\bf{C}}^{\scriptscriptstyle{(2)}}}{\partial_{\Phi} {\vec{H}^{*T}}}\right),
\end{equation}
where 
\begin{equation}
J=\vec{H}^{*T}\left({\bf M}^{\Phi\Phi}\partial_{\Phi}\vec{H}+{\bf B} \vec{H}\right).
\end{equation}
\noindent
Therefore, the flux number of gravitons associated with  rotationally invariant perturbations, i.e., the imaginary part of $J$, is given by
\noindent
\begin{equation}\label{Eq:ChargeBulk}
\mathcal{F}=-{\rm Im}\,J=\frac{i}{2}\left[\vec{H}^{*T}\left({\bf M}^{\Phi\Phi}\partial_{\Phi}\vec{H}+{\bf B} \vec{H}\right)-\left(\partial_{\Phi}\vec{H}^{*T} ({\bf M}^{\Phi\Phi})^{*T}+\vec{H}^{*T}{\bf B}^{*T}\right)\vec{H}\right].
\end{equation}
\noindent
Finally, this equation may be reduced by plugging Eqs.~\eqref{Eq:1}, \eqref{Eq:2} and \eqref{Eq:Matrices} in \eqref{Eq:ChargeBulk},
resulting in
\noindent
\begin{equation}
\mathcal{F}=-\frac{i \zeta_2 f}{3\,\zeta_1^2 (\partial_{\Phi}\zeta_1)^2}\left[H_{11}^{*}(\partial_{\Phi}H_{11})-(\partial_{\Phi}H_{11}^{*})H_{11}\right].
\end{equation}
\noindent
As observed in Ref.~\cite{Gubser:2008sz}, the flux number $F$ is proportional to the Wronskian of the complexified solution $H_{11}$. 
Therefore, in order to calculate $\mathcal{F}$, we only need the solution of $H_{11}$. 
Moreover, asymptotic solutions may be obtained considering the asymptotic behavior of the background. Close to the horizon, $H_{11}$ behaves like $H_{11}\propto f^{\alpha}$, where  $f$ is a function of the dilaton field $\Phi$. Plugging this ansatz into Eq.~\eqref{Eq:H11} we find two possible solutions:
\noindent
\begin{equation}\label{Eq:H11Horizon}
\alpha_1=\frac{i\omega\,\zeta_1(\Phi_h)}{\partial_{\Phi}f(\Phi_h)\,\zeta_2(\Phi_h)},\qquad \alpha_2=-\frac{i\omega\,\zeta_1(\Phi_h)}{\partial_{\Phi}f(\Phi_h)\,\zeta_2(\Phi_h)},
\end{equation}
\noindent
where $\Phi_h$ is the location of the horizon. In terms of the temperature, defined by
\noindent
\begin{equation}
T=-\frac{\partial_{\Phi}f(\Phi_h)}{4\pi}\frac{\zeta_2(\Phi_h)}{\zeta_1(\Phi_h)},
\end{equation}
\noindent
the first solution in \eqref{Eq:H11Horizon} represents waves falling into the black hole horizon, while the second solution represents waves coming out from the horizon. Thus, we choose $\alpha_1$ and we may write $H_{11}(\Phi)$ close to the horizon as
\noindent
\begin{equation}\label{Eq:BulkHorizon}
H_{11}=\mathcal{B}_{-}(\Phi_h-\Phi)^{-\frac{i\omega}{4\pi T}},
\end{equation}
\noindent
where $\mathcal{B}_{-}$ is a constant. Expanding the last function around $\omega=0$, we get
\begin{equation}
H_{11}\approx\mathcal{B}_{-}\left(1-\frac{i\omega}{4\pi T}\ln{(\Phi_h-\Phi)}\right).
\end{equation}
\noindent
Note that for $\omega=0$, the function $H_{11}$ becomes regular (constant) at the horizon. On the other hand, close to the boundary we consider the ansatz $H_{11}\propto \Phi^\beta$, 
plugging in Eq.~\eqref{Eq:3} we get two solution: $\beta_1=0$ and $\beta_2=(4-2\epsilon)/\epsilon$. Thus, the leading term in the asymptotic solution close to the boundary must be a constant, which can be fixed as
\noindent
\begin{equation}\label{Eq:BulkBoundary}
H_{11}(0)\to 1.
\end{equation}
\noindent

Now we are able to find  an expression for the flux number $\mathcal{F}$. Plugging \eqref{Eq:BulkHorizon} in \eqref{Eq:ChargeBulk} we get the simplified expression
\noindent
\begin{equation}
\mathcal{F}=\frac{2\omega}{3}\frac{|\mathcal{B}_{-}|^2}{\zeta_1(\Phi_h)\,(\partial_{\scriptscriptstyle{\Phi}}\zeta_1(\Phi_h))^2}.
\end{equation}
\noindent
Note that $\mathcal{F}$ depends on the (inverse) scale factor, its derivative, and the constant $|\mathcal{B}_{-}|$. The imaginary part of the retarded Green's function takes the form
\noindent
\begin{equation}
{\rm Im}\, G_{R}(\omega)=-\frac{\mathcal{F}}{16\pi G_5}=-\frac{\omega}{24\pi G_{5}}\frac{|\mathcal{B}_{-}|^2}{\zeta_1(\Phi_h)\,(\partial_{\scriptscriptstyle{\Phi}}\zeta_1(\Phi_h))^2}.
\end{equation}
\noindent
To calculate the bulk viscosity we follow the procedure of Ref.~\cite{Gubser:2008sz} and find that
\noindent
\begin{equation}\label{Eq:BulkViscosity}
\zeta=-\frac{4}{9}\lim_{\omega\to 0}\frac{1}{\omega}{\rm Im}\, G_{R}(\omega)=\frac{2}{27\pi}\frac{(\zeta_1(\Phi_h))^2}{(\partial_{\scriptscriptstyle{\Phi}}\zeta_1(\Phi_h))^2}\,s(\Phi_h)|\mathcal{B}_{-}|^2.
\end{equation}
\noindent
To write the last result we have used the entropy density relation \eqref{Eq:EntropyDensity}.

\begin{figure}[ht!]
\centering
\includegraphics[width=7cm]{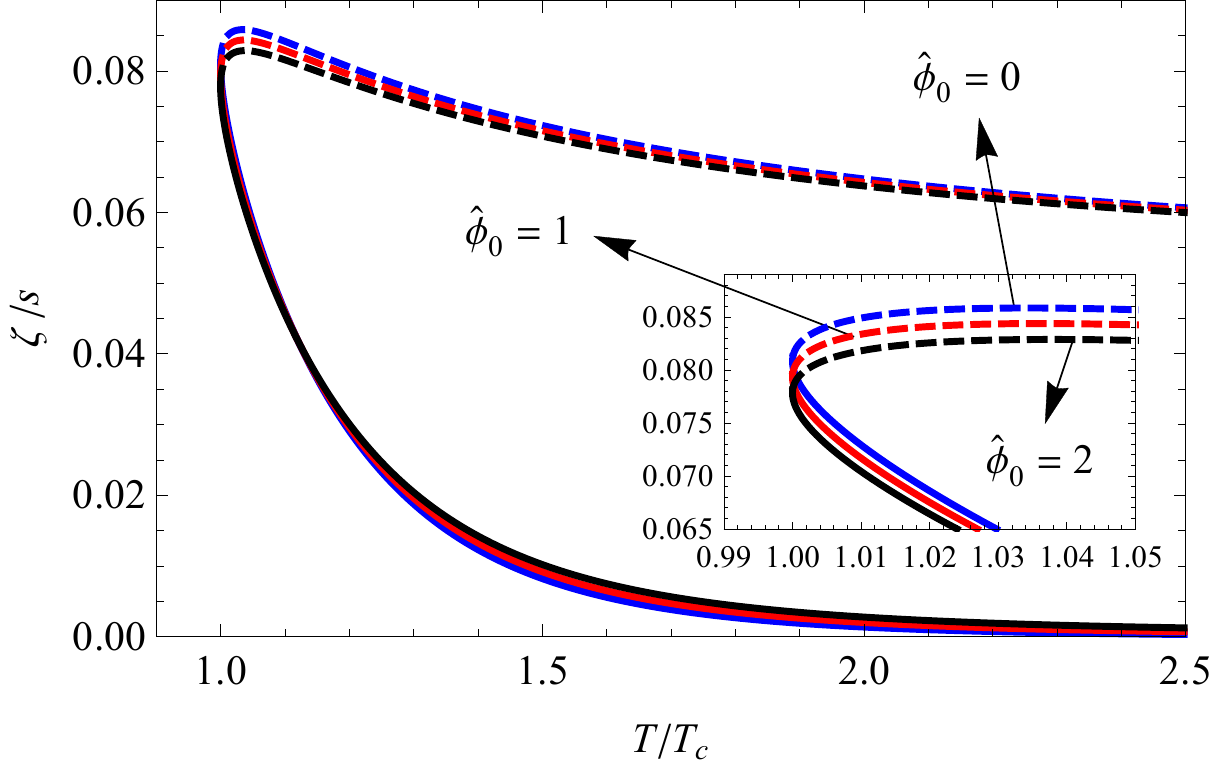}
\caption{
The bulk viscosity to entropy density ratio $\zeta/s$ as a function of $T/T_c$ for different values of $\hat\phi_0$ and $\epsilon=0.07$. Solid (dashed) lines represent the large (small) black hole regime. The results were obtained considering $|\mathcal{B}_{-}|=1$.
}
\label{Fig1:TBulk}
\end{figure}

A few comments are now in order. The bulk viscosity depends on the value of the inverse scale factor (and its derivative) evaluated at the horizon. Moreover, in contrast with the shear viscosity case, the bulk viscosity depends on the constant $|\mathcal{B}_{-}|$. 
It is worth mentioning that in the high-temperature regime the constant reduces to $|\mathcal{B}_{-}|=1$. Thus, in this region, the result \eqref{Eq:BulkViscosity} is in agreement with the formula obtained in Ref.~\cite{Eling:2011ms}. 
However, there is a subtlety when one compares both results, see the discussion in Ref.~\cite{Buchel:2011wx}. This is also true in the adiabatic approximation \cite{Eling:2011ms, Gursoy:2009kk} (see also \cite{Buchel:2011wx}). 
Considering $|\mathcal{B}_{-}|=1$, we display our numerical results for the bulk viscosity to entropy density ratio $\zeta/s$ in Fig.~\ref{Fig1:TBulk}. The solid lines represent the stable large black holes while dashed lines represent the unstable small black holes. 
We conclude that in the case $|\mathcal{B}_{-}|=1$ the ratio $\zeta/s$ is less sensitive to the value of $\hat\phi_0$. Note that the bulk viscosity has a sharp rise close to the critical temperature. This result was previously reported in the QCD literature, see for instance Ref.~\cite{Moore:2008ws} where a semi-analytic study is presented.

\begin{figure}[ht!]
\centering
\includegraphics[width=7cm]{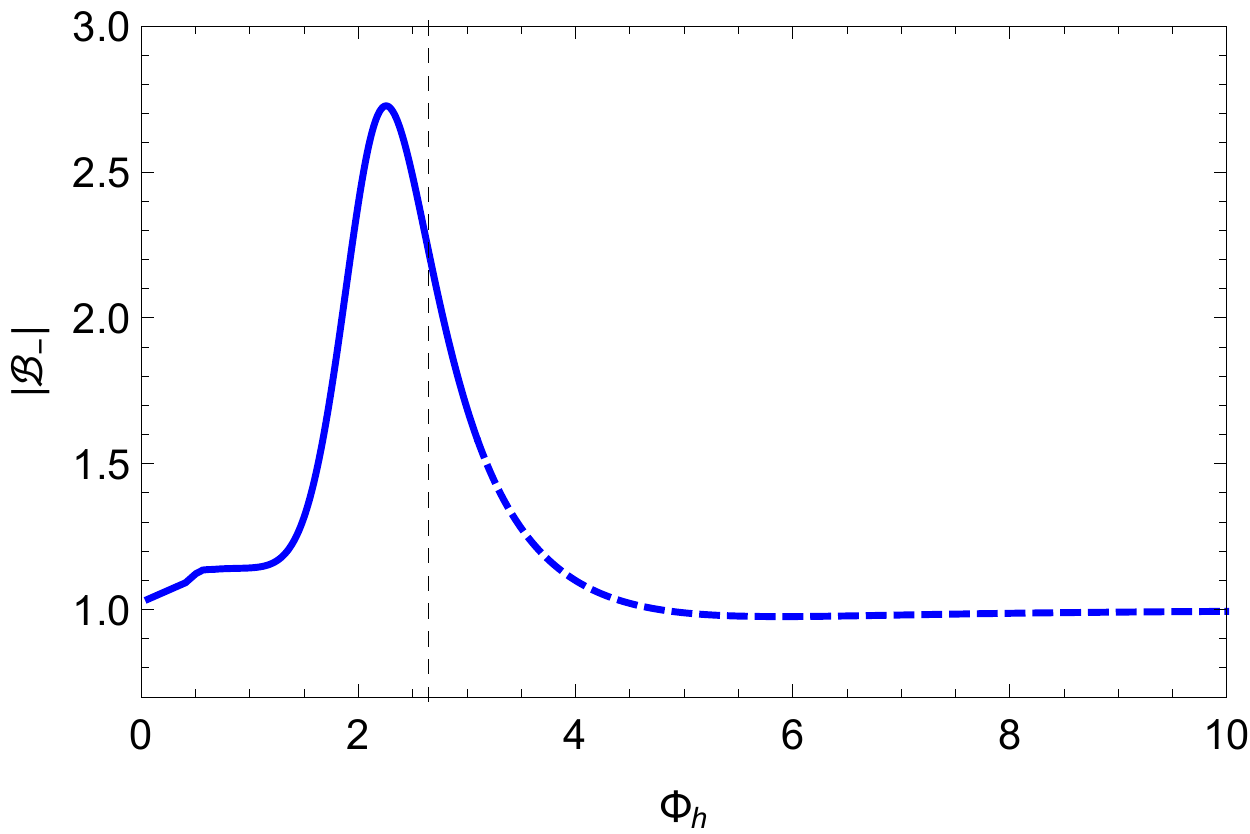}\hfill
\includegraphics[width=7cm]{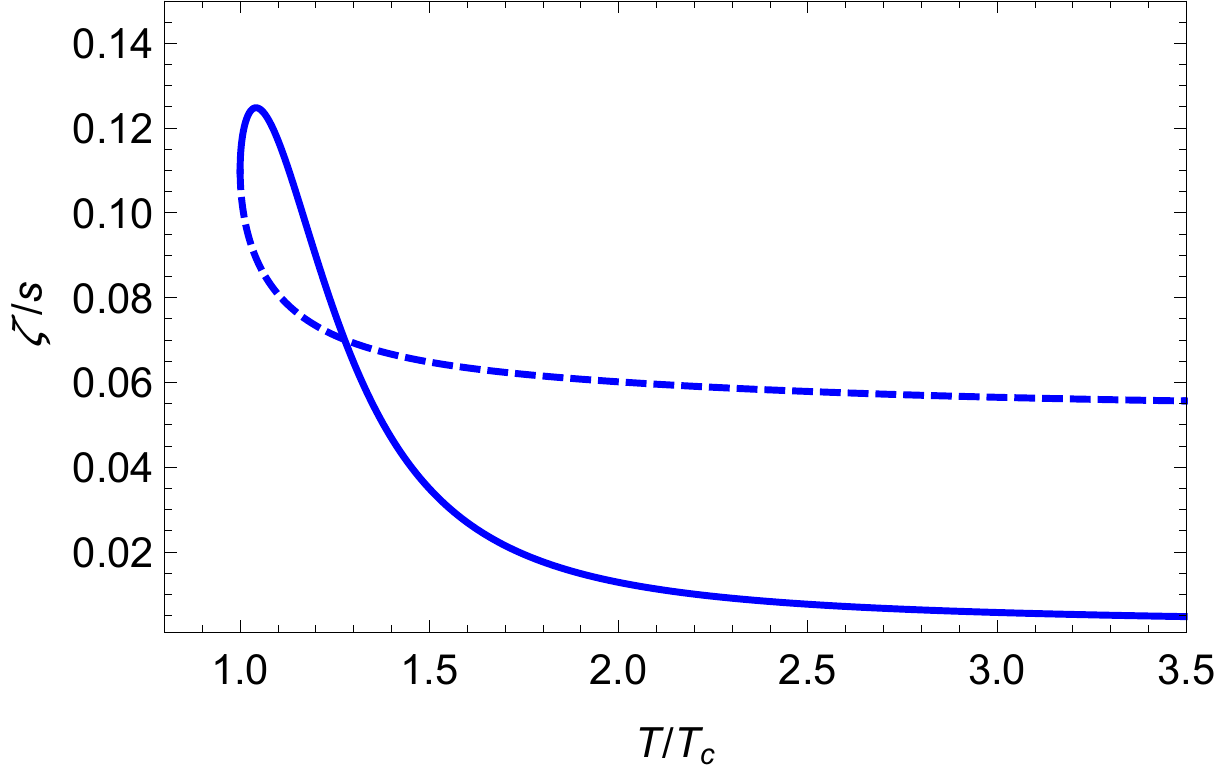}
\caption{
Left: The constant $|\mathcal{B}_{-}|$ as a function of $\Phi_h$ for $\hat\phi_0=2$, and $\epsilon=0.3$, vertical dashed line represents the place where the critical (minimal) temperature is reached. Right: Bulk viscosity to entropy density ratio as a function of $T/T_c$ calculated using Eq.~\eqref{Eq:BulkViscosity}.
}
\label{Fig1:TBulk2}
\end{figure}

On the other hand, we may calculate the constant $|\mathcal{B}_{-}|$ using the boundary conditions of the problem, the Dirichlet condition at the boundary $H_{11}(\Phi=0)=1$, and the incoming wave condition at the horizon given by \eqref{Eq:BulkHorizon}. Following this procedure, we calculate numerically $|\mathcal{B}_{-}|$ as a function of $\Phi_h$. The results are displayed on the left panel of Fig.~\ref{Fig1:TBulk2}. 
Meanwhile, the right panel of Fig.~\ref{Fig1:TBulk2} shows the bulk viscosity using $|\mathcal{B}_{-}|$ obtained numerically for $\hat\phi_0=2$ and $\epsilon=0.3$, solid lines represent the large black hole regime, while dashed lines small black hole regime. 
It is worth pointing out that depending on the combination of the parameters $\{\epsilon,\hat\phi_0\}$, we found values of $|\mathcal{B}_{-}|$ that does not satisfy the condition $|\mathcal{B}_{-}|\geq 1$. 
In fact, it is possible to get $|\mathcal{B}_{-}|=0$ in the region close to the critical temperature for the combination of parameters $\epsilon=0.1$ and $\hat\phi_0=2$, for example.

\begin{figure}[ht!]
\centering
\includegraphics[width=7cm]{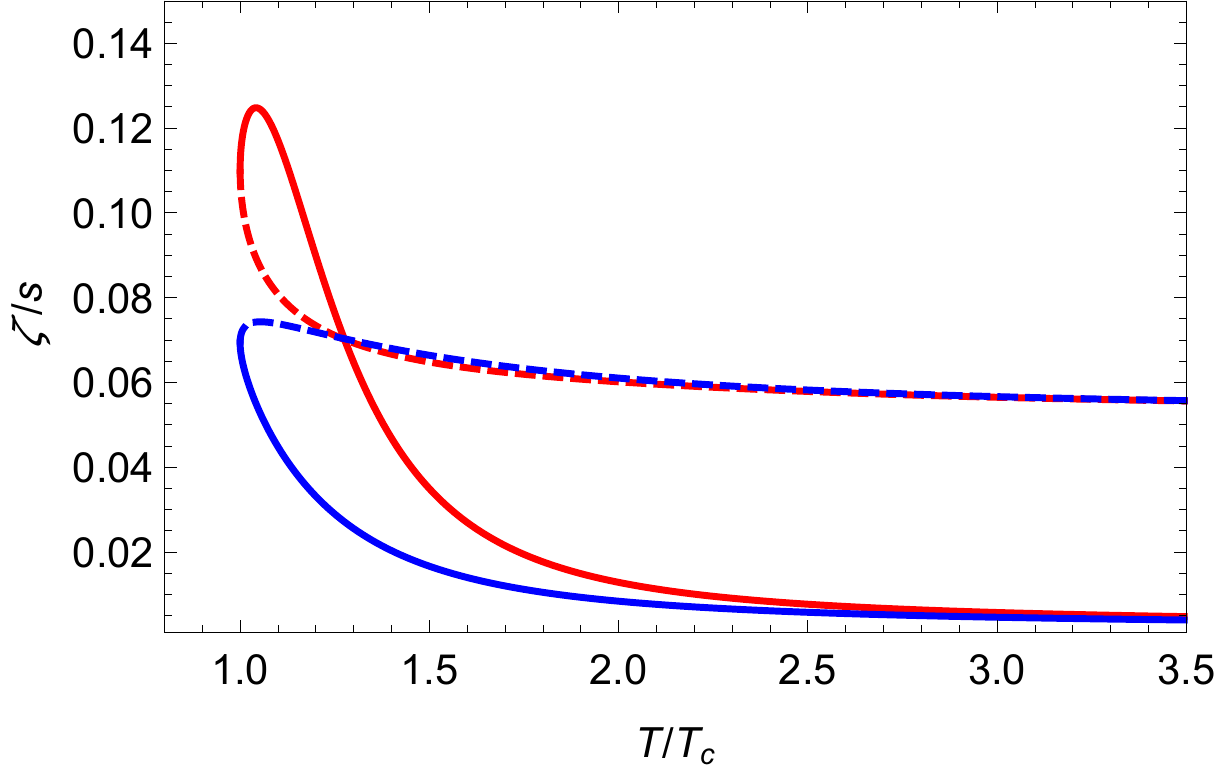}
\caption{
The figure shows the bulk viscosity to entropy density ratio as a function of $T/T_c$ for $\epsilon=0.3$, and $\hat\phi_0=2$. The result of Eq.~\eqref{Eq:BulkViscosity} is displayed with a red line, while the result considering $|\mathcal{B}_-|=1$ with a blue line.
}
\label{Fig1:TBulkGPREO}
\end{figure}

Knowing the asymptotic form of the inverse scale factor we are able to find an expression for the bulk viscosity to entropy density ratio in the regime of large black holes. 
In this regime the derivative of the horizon function may be neglected, this means that the coefficient of $H_{11}(\Phi)$ in Eq.~\eqref{Eq:3Improved} is zero in the limit of zero frequency. Thus, we may solve the resulting differential equation getting the asymptotic solution,
\noindent
\begin{equation}
H_{11}(\Phi)=c_1+c_2\,\Phi^{\frac{4-2\epsilon}{\epsilon}}.
\end{equation}
Hence, the boundary condition $H_{11}(0)\to 1$ reduces the solution to $c_1=|\mathcal{B}_{-}|=1$. Plugging this result and the leading term of the asymptotic warp factor \eqref{asymptotics2} into \eqref{Eq:BulkViscosity}, we get
\noindent
\begin{equation}
\frac{\zeta}{s}=\frac{2(\epsilon\,\hat\phi_0)^2}{27\pi}\frac{\Lambda^{2\epsilon}}{\left(\pi T\right)^{2\,\epsilon}}.
\quad  \text{(High temperature)}
\end{equation}
\noindent
From this expression, it is easy to see the role of the parameter $\Lambda$. Notice that the bulk viscosity vanishes in the limit $\Lambda \to 0$ where the dilaton field \eqref{Eq:DilModelA1} vanishes, recovering conformal symmetry. We obtained numerical results for \eqref{Eq:BulkViscosity} considering $|\mathcal{B}_-|=1$ and $|\mathcal{B}_-|$ 
obtained numerically, an overlap of both results is displayed in Fig.~\ref{Fig1:TBulkGPREO}, where the blue line represents the case $|\mathcal{B}_-|=1$, while the red line represents the case where $|\mathcal{B}_-|$ is obtained numerically. 
As it can be seen from the figure, both results are in agreement in the region of high-temperatures for both the large black holes (solid lines) and small black holes (dashed lines). 

\section{Comparison with lattice $SU(N_c)$ theories and heavy ion collisions}
\label{Sec:Data}

In this section we compare our results for the thermodynamics quantities against the results from lattice $SU(N_c)$ theories \cite{Panero:2009tv, Lucini:2012wq, Lucini:2013qja} and we also compare our results for the viscosity coefficients against the results from the JETSCAPE collaboration \cite{Everett:2020yty}, 
obtained from a model-to-data analysis of heavy-ion collision experimental data. For this comparison, we will only be interested in the (physical) stable black hole solution, namely the large black hole. 

\subsection{Thermodynamic quantities}
\label{SubSec:Thermoqts}

Let us start this subsection by describing the critical temperature for the formation of a non-conformal plasma in our EHQCD model. In our approach, we identify the critical temperature with the minimum temperature for the existence of black hole solutions. 

\begin{figure}[ht]
\centering
\includegraphics[width=7cm]{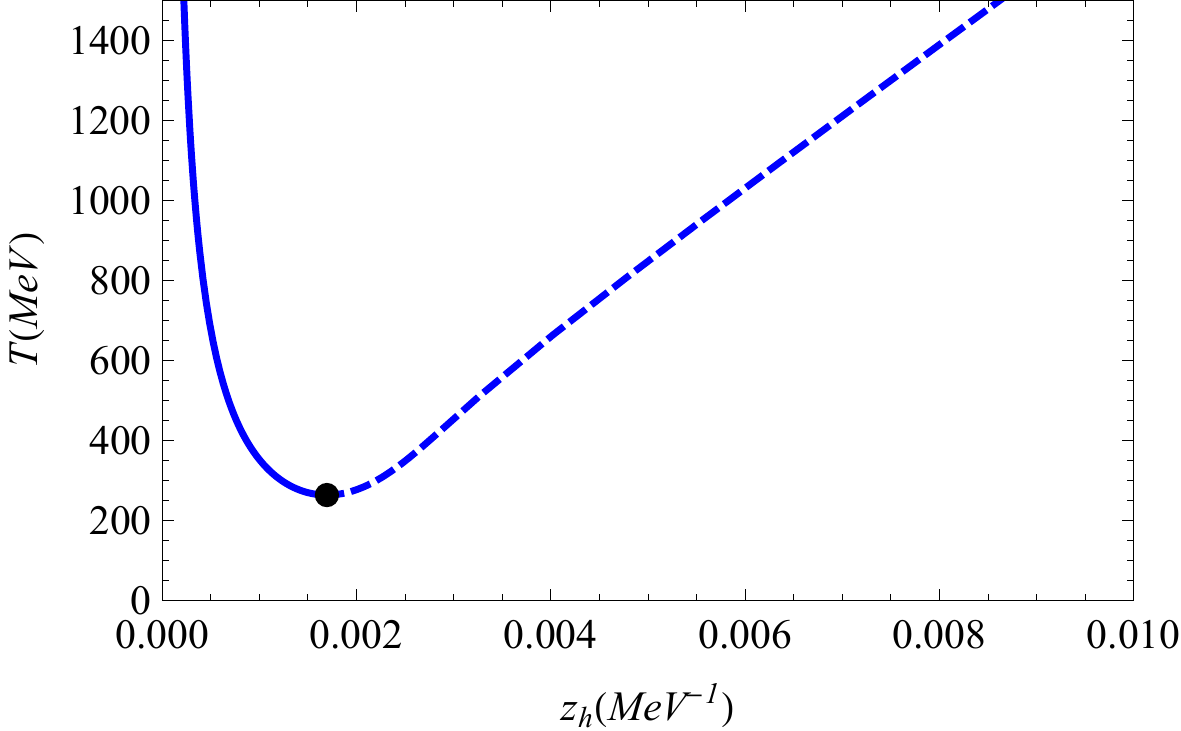}\hfill
\includegraphics[width=7cm]{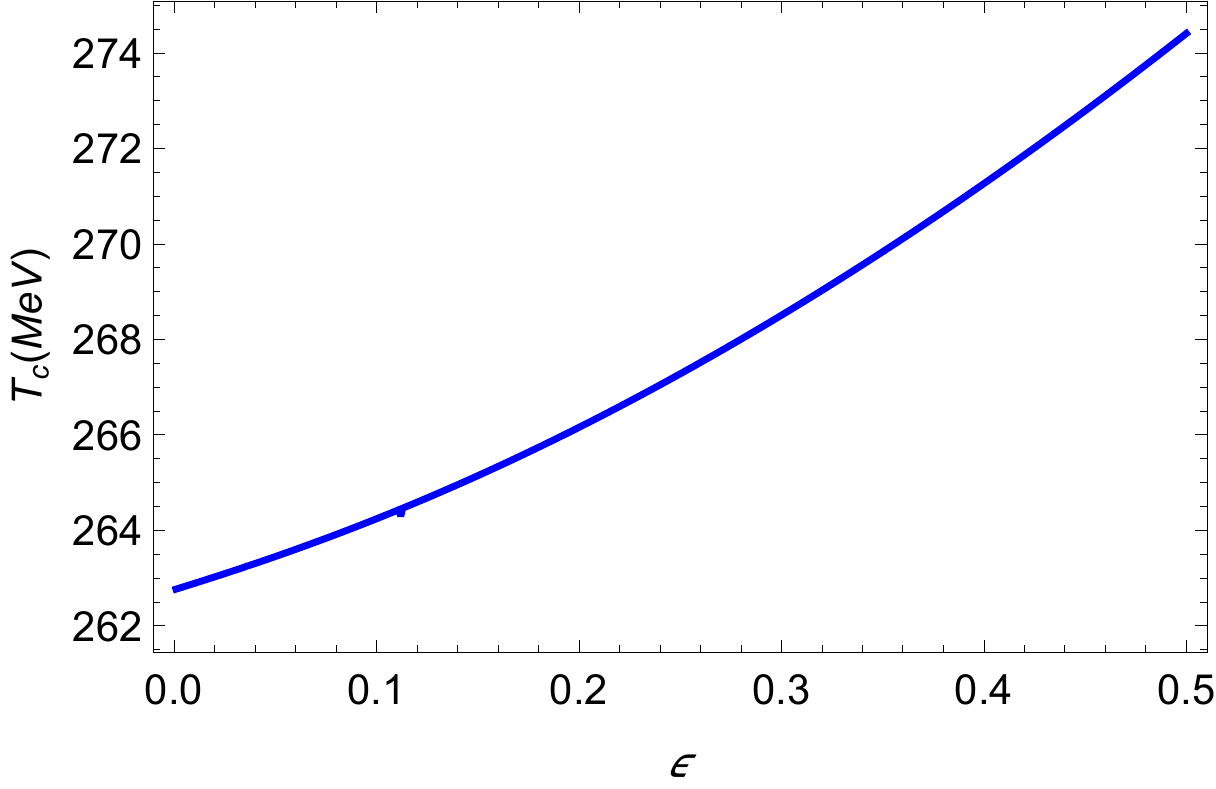}
\caption{
Left: The temperature as a function of $z_h$ for $\hat\phi_0=5.6$ and $\epsilon=0.1$. The black dot shows the location of the minimum. Right: The critical temperature as a function of the conformal anomalous dimension.
}
\label{Fig1:EpsilonTFAP}
\end{figure}

Considering the parameters fixed by the glueball spectrum, obtained in Ref.~\cite{Ballon-Bayona:2017sxa}, we calculate the critical temperature and observe its dependence on the conformal anomalous dimension $\epsilon$. 
The critical temperature is obtained by solving the equation $\partial_{z_h}T=0$, and it is represented by a black dot in the plot displayed on the left panel of Fig.~\ref{Fig1:EpsilonTFAP}. 
The plot describes the variation of the temperature (in MeV) as a function of $z_h$,  where $\hat \phi_0$ and $\Lambda$  (in MeV) were fixed for a given $\epsilon$ from a fit to the glueball spectrum, as done previously in Ref.~\cite{Ballon-Bayona:2017sxa}. 
For the particular case shown in the figure, we have  $\epsilon=0.1$, $\hat\phi_0=5.6$ and $\Lambda=743 \, {\rm MeV}$. As described in the previous sections, there are two branches: one corresponds to the large black hole (solid line) and the other corresponds to the small black hole (dashed line). 

In turn, the right panel of Fig.~\ref{Fig1:EpsilonTFAP} shows the critical temperature (in MeV) as a function of the conformal anomalous dimension $\epsilon$. For each value of $\epsilon$ the parameters $\hat \phi_0$ and $\Lambda$ (in MeV) were fixed appropriately from the glueball spectrum \cite{Ballon-Bayona:2017sxa}. 
The interval considered here is $\epsilon\in [10^{-3},0.5]$, extending the fit to the glueball spectrum performed in \cite{Ballon-Bayona:2017sxa}. As the figure shows, the critical temperature is a slowly growing function of the conformal anomalous dimension $\epsilon$ and varies from  $T_c\approx 263\,\text{MeV}$ to $T_c \approx 274 \,\text{MeV}$. 
It is worth pointing out that the present results for the critical temperature are very close to the recent results found in lattice $SU(N_c)$  theories \cite{Lucini:2012wq, Lucini:2013qja}, namely $T_c \approx 0.595 \sqrt{\sigma} \approx 262\,\text{MeV}$ where $\sigma \approx (440 \, {\rm MeV})^2$ is the phenomenological value for the string tension. 
It is remarkable that a holographic model with the metric coupled to a single scalar field provides such a result. This analysis also suggests an alternative approach for fixing the parameter $\Lambda$. 
We will see later in this section that for a given value of the anomalous conformal dimension $\epsilon$ we can use the lattice results for the trace anomaly to fix the dimensionless coupling $\hat \phi_0$. Then we can also use the lattice result for $T_c$ to fix $\Lambda$ in MeV units. 
In conclusion, the parameters $\hat \phi_0$ and $\Lambda$ can be obtained as a function of $\epsilon$ considering either the glueball spectrum at zero temperature, as done in Ref.~\cite{Ballon-Bayona:2017sxa}, or the thermodynamics of the non-conformal plasma at finite temperature. 

\begin{figure}[ht]
\centering
\includegraphics[width=7cm]{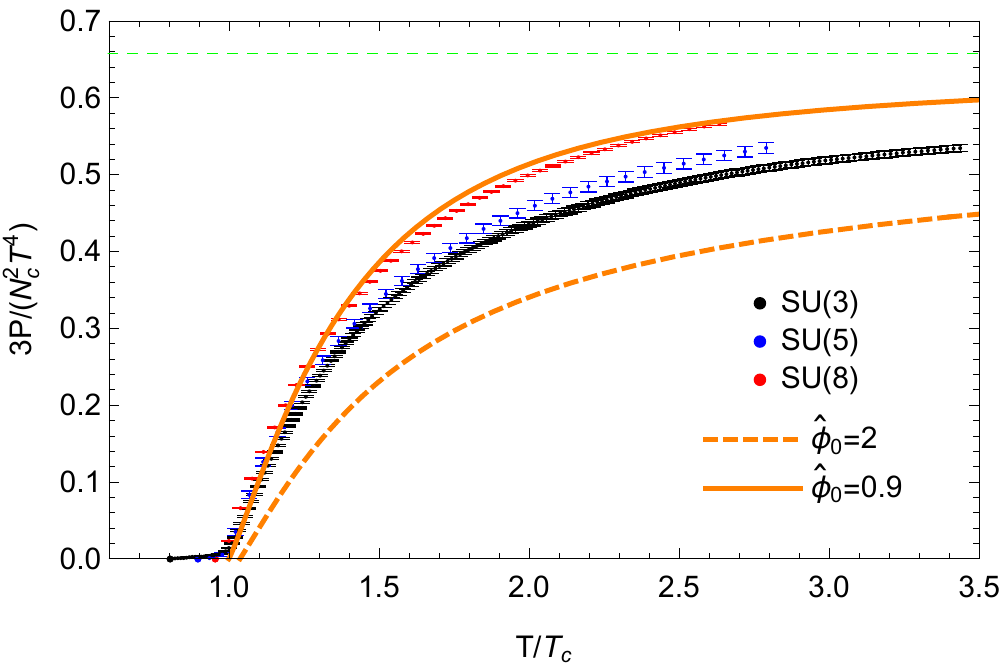}\hfill
\includegraphics[width=7cm]{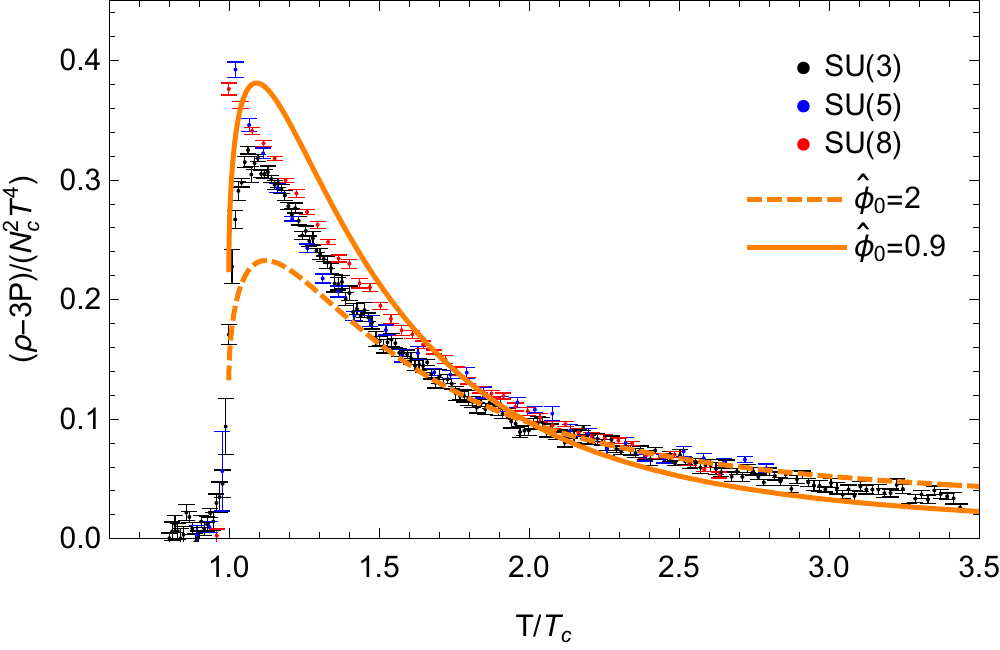}
\caption{
Left: The normalized pressure as a function of $T/T_c$.   Right: The  normalized trace anomaly as a function of $T/T_c$.  The lattice results (discrete/dotted lines) include error bars, while the results of the holographic model are represented by orange lines.  Both figures were obtained for $\epsilon=0.3$.
}
\label{Fig:TraceAnomaly}
\end{figure}

Next, we compare our results for the pressure of the non-conformal plasma against the results obtained in lattice $SU(N_c)$ theories \cite{Panero:2009tv}. 
Our results for the pressure, properly normalized, as a function of $T/T_c$ are displayed in the left panel of Fig.~\ref{Fig:TraceAnomaly} (solid lines) and compared against the lattice results  (dotted lines with error bars). As the figure shows, the pressure in our EHQCD model is sensitive to the value of the parameter $\hat\phi_0$, once the parameter $\epsilon$ is kept fixed. 
We present results for two values of $\hat\phi_0$: one of them being fixed with the glueball spectrum at zero temperature, as in \cite{Ballon-Bayona:2017sxa}, and the other being fixed by matching the maximum value for the dimensionless trace anomaly  $(\rho-3p)/(N_c^2T^4)$ with the lattice result $0.381$ found for $SU(8)$ \cite{Panero:2009tv}. 
Although displaying a qualitative agreement with the lattice results, we observe that the values of the pressure in the case of parameters fixed by the glueball spectrum, i.e., for $\{\epsilon,\hat\phi_0\}=\{0.3,2\}$ (orange dashed line), are far from a quantitative agreement.
This apparent shortcoming was also observed in other holographic models for QCD, see for instance \cite{Gursoy:2008bu}. 
In turn, the results for the pressure in the case the parameters were fixed by the trace anomaly condition, i.e., for $\{\epsilon,\hat\phi_0\}=\{0.3,0.9\}$ (orange solid line), are in quantitative agreement with those obtained from QCD on the lattice in the limit of large $N_c$.

It is also very interesting to compare the trace anomaly against the lattice $SU(N_c)$ results. Our results for the trace anomaly, properly normalized, as a function of $T/T_c$, are displayed in the right panel of Fig.~\ref{Fig:TraceAnomaly} (orange lines) and compared against the lattice results  \cite{Panero:2009tv} (dotted lines with error bars). 
The figure shows that the results provided by the combination of parameters $\{\epsilon,\hat\phi_0\}=\{0.3, 0.9\}$ (orange solid line) are in quantitative agreement with those of the lattice theory. 
In turn, the set of parameters $\{\epsilon,\hat\phi_0\}=\{0.3,2\}$ (orange dashed line), consistent with the glueballs spectrum, are not in quantitative agreement with those obtained on the lattice. 

We have found in this section that our EHQCD model with parameters $\hat \phi_0$ and $\Lambda$ fixed as a function of $\epsilon$ from a fit to the glueball spectrum provides at finite temperature a good result for the critical temperature for deconfinement and pressure that displays a qualitative agreement with lattice $SU(N_c)$ theories. 
However, for the pressure we could not obtain a quantitative agreement; the pressure in our EHQCD model is lower than the lattice result when the parameters are fixed by the (zero temperature) glueball spectrum. We also found a trace anomaly that has a qualitative behavior similar to that obtained in lattice $SU(N_c)$ theories but presents a peak that is lower than the corresponding lattice result. 
Since the main goal of holographic QCD models is to provide an effective description of QCD in the limit of large $N_c$, the fact that we found a discrepancy with the lattice $SU(N_c)$ results for the thermodynamic quantities suggests that some more ingredients may be needed in our EHQCD approach.

\begin{figure}[ht!]
\centering
\includegraphics[width=7cm]{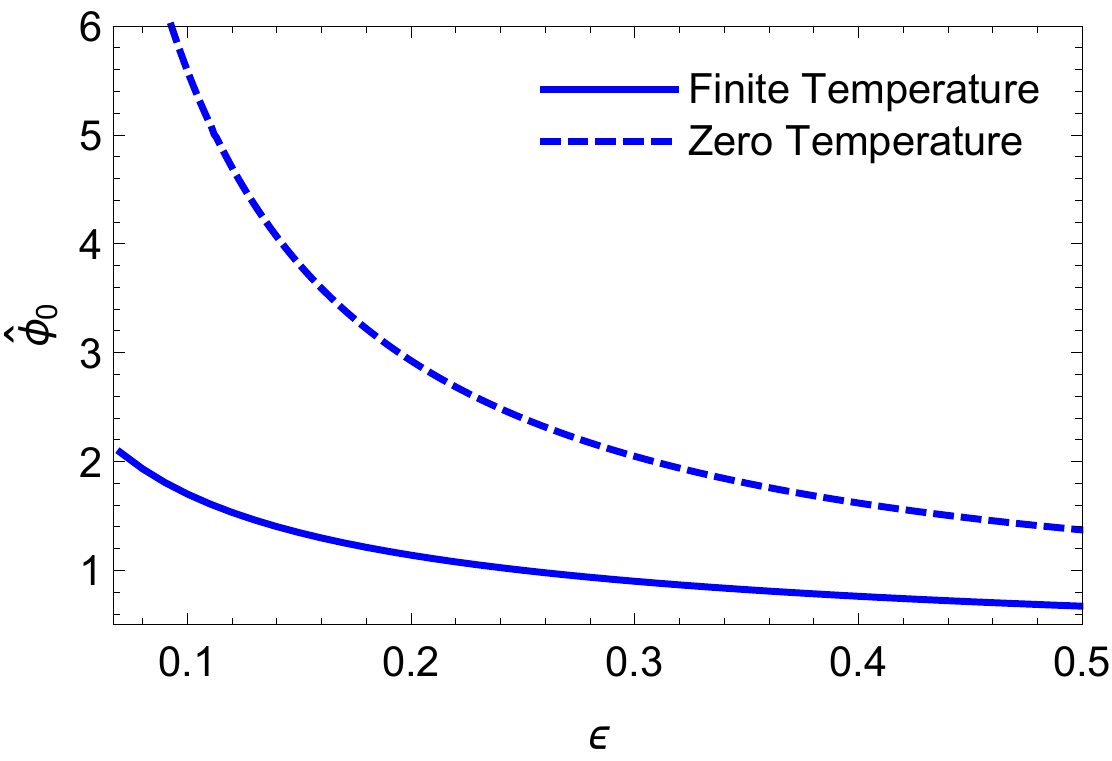}\hfill
\includegraphics[width=7cm]{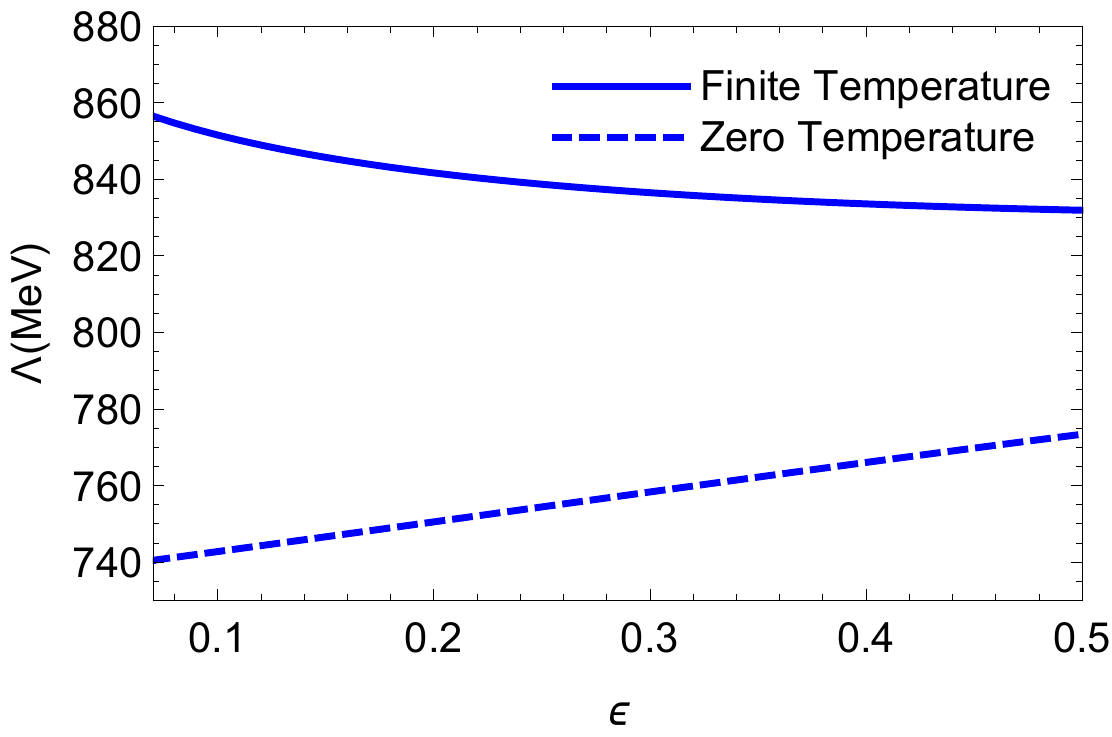}
\caption{
The behavior of $\hat\phi_0$ (left panel) and $\Lambda$ (right panel) as a function of $\epsilon$ with the other parameters fixed by the available data from $SU(8)$ theory on the lattice is shown by the solid blue lines, in comparison to the results obtained when the other parameters are fixed by the glueball spectrum at zero temperature, represented by the dashed blue lines. 
}
\label{Fig1:Phi0vsEp}
\end{figure}

A phenomenological solution for the discrepancy was considered in Ref.~\cite{Gursoy:2009jd}, where additional parameters were added to control the curve of the pressure in the region of high temperatures and the height of the peak of the trace anomaly close to the critical temperature. 
Our analysis suggests the following alternative approach. Once the parameter $\epsilon$ was fixed, we can fix the dimensionless coupling $\hat\phi_0$ imposing the condition $(\rho-3p)/(N_c^2T^4)=0.381$, which corresponds to the maximum value of the dimensionless trace anomaly obtained in the lattice $SU(8)$ theory \cite{Panero:2009tv}. 
The running of $\hat\phi_0$ as a function of $\epsilon$ is displayed on the left panel of Fig.~\ref{Fig1:Phi0vsEp} (solid line) and compared to the case where $\hat\phi_0$ was fixed from a fit to the glueball spectrum (dashed lined). In turn, the parameter $\Lambda$ can be fixed by matching the critical temperature for deconfinement $T_c$ with the value $262 {\rm MeV}$ obtained in lattice $SU(N_c)$ theories in the limit of large $N_c$. 
The dependence of $\Lambda$ on $\epsilon$ is displayed on the right panel of Fig.~\ref{Fig1:Phi0vsEp} (solid line) and compared to the case where $\Lambda$ was fixed from a fit to the glueball spectrum (dashed line). 

We therefore reach the following conclusion. In our EHQCD model, when one fixes the conformal anomalous dimension $\epsilon$ there are two possible methods for fixing the parameters $\hat \phi_0$ and $\Lambda$. The first method uses a fit to the lattice results for the glueball spectrum at zero temperature to fix $\hat \phi_0$ and $\Lambda$. 
This is the method we followed in our previous work \cite{Ballon-Bayona:2017sxa}. The parameters fixed in that way are displayed as dashed lines in Fig.~\ref{Fig1:Phi0vsEp}. The second method, found in this work, uses the results for some thermodynamics quantities describing the gluon plasma in lattice $SU(N_c)$ theories to fix $\hat \phi_0$ and $\Lambda$, 
namely the maximum for the dimensionless trace anomaly to fix  $\hat \phi_0$ and the value for the deconfinement temperature to fix $\Lambda$. The parameters fixed by following the second method are displayed as solid lines in Fig.~\ref{Fig1:Phi0vsEp}.

Note that the behavior of $\hat\phi_0(\epsilon)$ is qualitatively similar for both methods (solid and dashed lines). As regards the behavior of $\Lambda(\epsilon)$, in the first method we find that $\Lambda$ is a slowly growing function of $\epsilon$ whereas the second method yields 
$\Lambda$ as a slowly decreasing function of $\epsilon$. On physical grounds, since the parameter $\Lambda$ represents a dynamical mass gap similar to $\Lambda_{QCD}$ it is expected that it should be independent of the parameter $\epsilon$. 
The fact that we find a slow variation for $\Lambda(\epsilon)$ suggests that some further improvement would allow us to satisfy this physical condition.

We would like to remark that the numerical errors for the thermodynamic quantities in our model are very small. 
This is because the only differential equation that is solved numerically is the first equation in \eqref{EqsBH} that allows us to find the (inverse) scale factor $\zeta_1$ as a function of $z$ for a given dilaton profile $\Phi(z)$. 
Considering a moderate numerical precision,  the uncertainty in $\zeta_1$ is of order $10^{-6}$. The entropy density is inversely proportional to $\zeta_1(z_h)^3$ and therefore has an uncertainty of the same order. All the other thermodynamic quantities were obtained from the entropy density.

\subsection{Shear and bulk viscosity}
\label{SubSec:ShearBulk}

\begin{figure}[ht!]
\centering
\includegraphics[width=7cm]{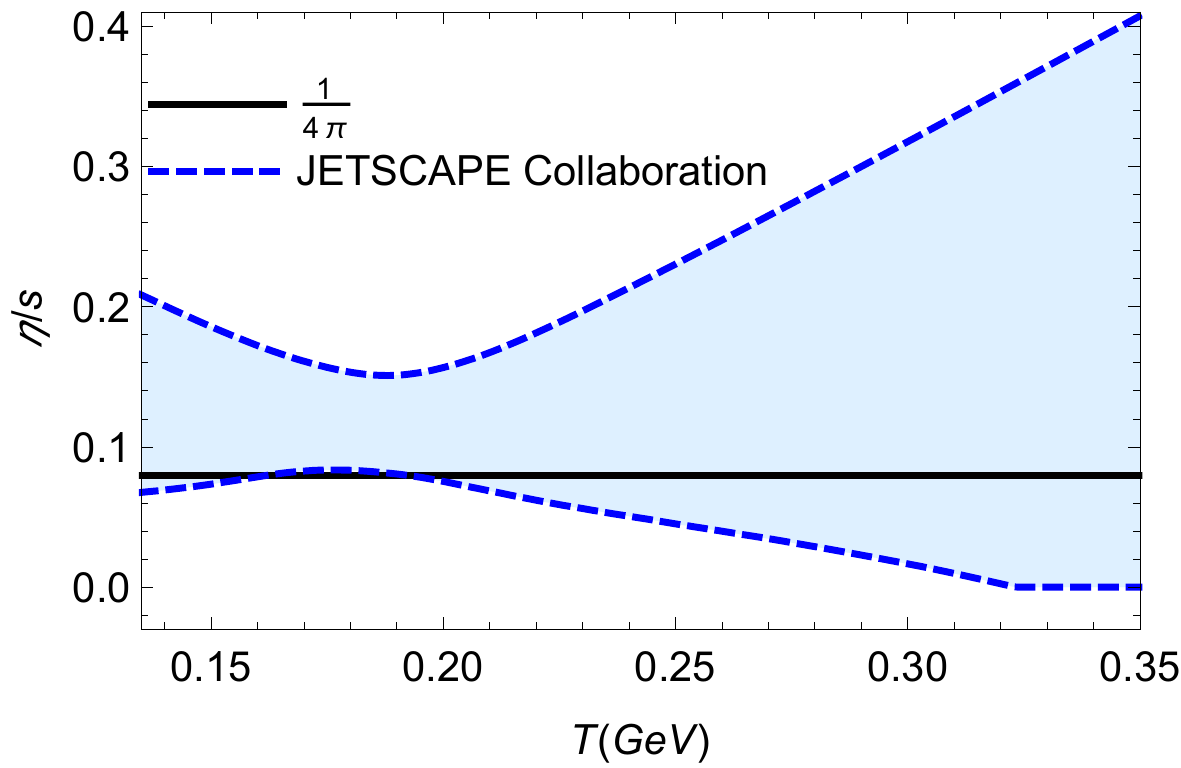}\hfill
\includegraphics[width=7cm]{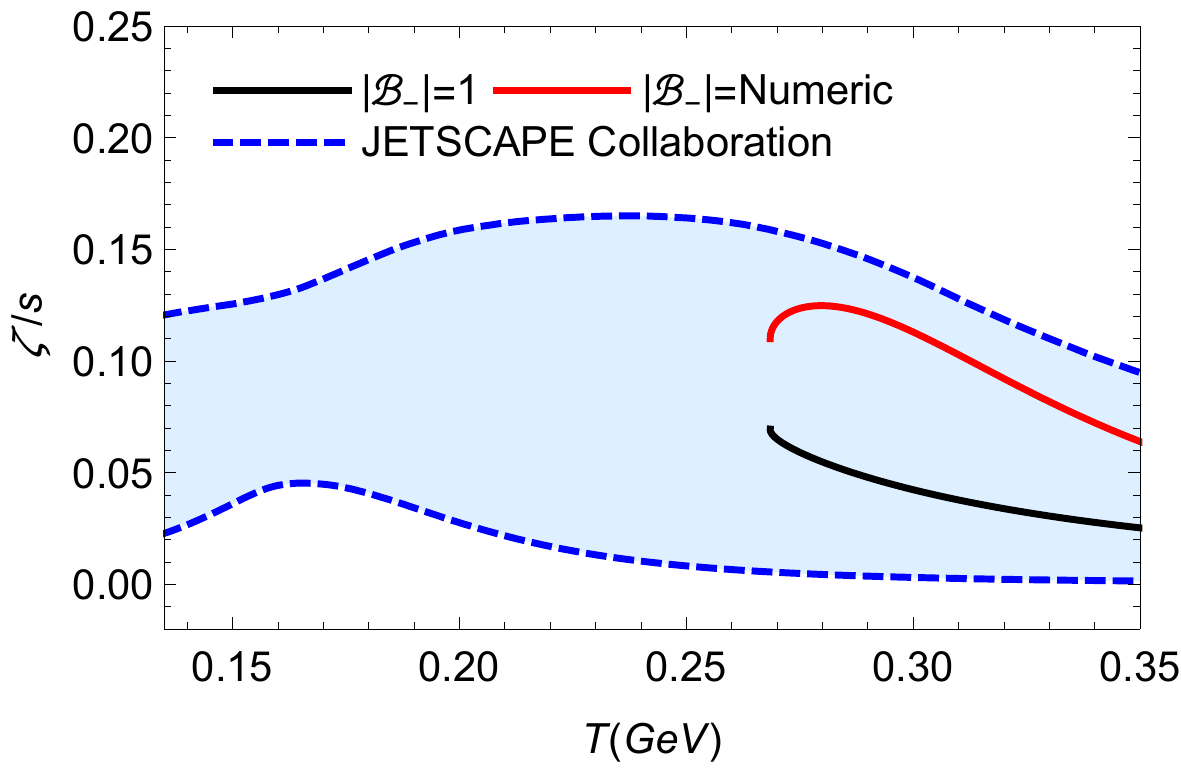}
\caption{
Left: Shear viscosity to entropy density ratio as a function of the temperature. The black solid line represents the holographic result. The region bounded by dashed blue lines corresponds to the interval constrained by the JETSCAPE collaboration \cite{Everett:2020yty}. Right: Bulk viscosity to entropy density ratio $\zeta/s$ as a function of $T$ for $\hat\phi_0=2$, and $\epsilon=0.3$. The black line was drawn by taking $|\mathcal{B}_-|=1$. The red line represents the results for $\zeta/s$ by taking the value of $|\mathcal{B}_-|$ determined numerically from the corresponding critical temperature $T_c=0.269\,\text{GeV}$.
}
\label{Fig1:TBulkComp}
\end{figure}

We end this section by comparing our results for the viscosity coefficients in the regime of large black holes against the results obtained by the JETSCAPE collaboration \cite{Everett:2020yty}  (see also \cite{Everett:2020xug}), 
from a model-to-data analysis of by e heavy-ion collision data. The universal result for the shear to entropy density ratio $\eta/s$ reproduced in this work, is displayed by the solid black line on the left panel of Fig.~\ref{Fig1:TBulkComp}, while the dashed blue lines represent the upper and lower bounds found by the JETSCAPE collaboration. 
As it can be seen from the figure, the universal result  $\eta/s=1/(4\pi)$ fits well within the region bounded by the lines drawn from the experimental data, except in the small interval of temperatures $0.15\text{---}0.20\,\text{GeV}$ where the solid line lies below the lower dashed line. 

The bulk viscosity to entropy density ratio $\zeta/s$ is displayed on the right panel of Fig.~\ref{Fig1:TBulkComp}. The sharp rise close to the critical temperature is in agreement with previous calculations on the lattice \cite{Meyer:2007dy} (see also \cite{Kharzeev:2007wb}). 
Using Eq.~\eqref{Eq:BulkViscosity} with $|\mathcal{B}_-|=1$, the maximum value of the ratio $\zeta/s\approx 0.0695$ (black line). Meanwhile, considering $|\mathcal{B}_-|$ obtained numerically, the maximum value of the ratio is $\zeta/s\approx 0.125$ (red line), for the set of parameters $\{\epsilon,\hat\phi_0, \Lambda\}=\{0.3,2,759 \, {\rm MeV}\}$ fixed by a fit to the glueball spectrum. 
Compared against the lattice results of \cite{Meyer:2007dy}, our results are within the error bars. In turn, our results are a bit larger than the results of \cite{Kharzeev:2007wb}. However, our results fit very well into the expected region presented recently by the JETSCAPE Collaboration \cite{Everett:2020yty, Everett:2020xug} enclosed by dashed blue lines. 
Compared against the holographic models of Refs.~\cite{Gubser:2008sz,Gursoy:2009kk}, the result for $|\mathcal{B}_-|=1$ is of the same order, but for $|\mathcal{B}_-|$ obtained numerically our result is larger than the one obtained in these papers, 
see also Refs.~\cite{Gursoy:2009kk, Finazzo:2014cna, Yaresko:2013tia, Li:2014dsa} for additional discussions. Hence, we conclude that our results are in agreement with those results obtained previously in the literature.

We would like to remark that our result for the shear viscosity to entropy ratio $\eta/s=1/(4\pi)$ is exact whereas our result for the bulk viscosity to entropy ratio $\zeta/s$ is obtained numerically using the formula \eqref{Eq:BulkViscosity}. 
According to this formula, the bulk to entropy ratio  $\zeta/s$ only depends on powers of the inverse scale factor $\zeta_1$, its derivative, and the coefficient ${\cal B}_{-}$. 
For a moderate numerical precision, we find that the uncertainty in the scale factor $\zeta_1$ is of order $10^{-6}$ and the uncertainty in the coefficient ${\cal B}_{-}$ is of order $10^{-4}$. 
Thus, the uncertainty in $\zeta/s$ is approximately of order $10^{-4}$. This uncertainty is very small compared with the uncertainties found in lattice QCD.

\section{Conclusions}
\label{Sec:Conclusions}

We have described the finite-temperature extension of the effective holographic models for QCD (EHQCD) proposed in \cite{Ballon-Bayona:2017sxa} in order to describe the physics of a non-conformal plasma. 
At zero temperature the EHQCD model provides a spectrum of scalar and tensor glueballs in agreement with results obtained in lattice QCD \cite{Meyer:2004gx}. The finite-temperature extension consists in embedding a black hole solution into the gravitational Einstein-dilaton theory, which is equivalent to creating a thermal state in the dual quantum field theory. 
We calculated some of the relevant thermodynamic variables, which are required to investigate the stability of the black holes. We showed that the large black holes are thermally stable, while the small black holes are thermally unstable. 
The unstable black hole is characterized by a negative specific heat and an imaginary speed of sound. We also showed that all the relevant thermodynamic quantities are sensitive to the variation of the model parameters, namely the dimensionless coupling $\hat\phi_0$ and the conformal anomalous dimension $\epsilon$. 
These two parameters characterize the breaking of conformal symmetry in EHQCD. Interestingly, we found that the pressure and the trace anomaly display qualitative behaviors that are similar to the ones found in lattice $SU(N_c)$ theories. 
In particular, the trace anomaly displays a peak near the critical temperature for deconfinement.  In the limit of very high temperatures, all thermodynamic quantities reach the corresponding conformal values. 
 
We also investigated the viscosity coefficients associated with the transport properties of the non-conformal fluid. We recovered the universal result for the ratio between the shear viscosity and the entropy density $\eta/s=1/(4\pi)$. 
This result was found by writing the relevant equations for the metric perturbations in terms of a gauge-invariant variable, calculating the retarded Green's function, and extracting the shear viscosity through Kubo's formula. 
Our numerical results show that the shear viscosity and the entropy density behave almost identically with the temperature, rising rapidly close to the critical temperature and then varying slowly in the region of high temperatures. 
We verified that the shear viscosity is sensitive to the variation of $\hat\phi_0$ for fixed $\epsilon$, and vice-versa, while the ratio $\eta/s$ remains constant, as expected for this class of holographic models arising from Einstein-dilaton gravity. 
To study the bulk viscosity we followed a different approach, by adopting a gauge where the dilaton field plays the role of the holographic coordinate. After calculating the retarded Green's function we were able to find an expression for the bulk viscosity by using Kubo's formula. 
The numerical results indicate that the holographic bulk viscosity increases sharply close to the critical temperature. This result is in agreement with previous predictions from lattice QCD and other holographic models of QCD. 
We also showed that the bulk viscosity is sensitive to the parameters of the model $\hat\phi_0$ and $\epsilon$. It is worth mentioning that our results were obtained using Model A1 proposed in Ref.~\cite{Ballon-Bayona:2017sxa}. However, we have also considered Model A2 proposed in that work and obtained equivalent results. 
We decided not to present the results of Model A2 in this paper to avoid redundancy. We believe that Models B1 and B2 would provide equivalent results to those obtained from Models A1 and A2.

Finally, we compared our results on the thermodynamics against the data available from lattice $SU(N_c)$ theories. We also compared our results for the viscosity coefficients against those found by the JETSCAPE collaboration. 
Regarding thermodynamic quantities, we found that our results for the pressure and trace anomaly are in qualitative agreement with the results found in the literature for lattice $SU(N_c)$ theories. 
However, we found that the value of the parameters adjusted to fit the glueball spectrum does not provide a quantitative agreement with the thermodynamics of lattice $SU(N_c)$ theories. 
We found, however, that a quantitative agreement with lattice $SU(N_c)$ theories is possible if one fixes the model parameters $\hat \phi_0$ and $\Lambda$, for a given value of $\epsilon$, by using the lattice results for the maximum value of the dimensionless trace anomaly and the critical temperature for deconfinement, respectively. 
We concluded that the results for the viscosity coefficients provided by our EHQCD model are consistent with the phenomenological constraints obtained by the JETSCAPE collaboration from the model-to-data analysis of the heavy-ion collision data. 
Although the shear viscosity did not always belong to the region bounded by the JETSCAPE collaboration, the bulk viscosity fits very well in the region of parameters considered by the collaboration. 

A possible  extension of this work would be by including flavor degrees of freedom in order to investigate chiral symmetry breaking. This task shall be reached by adding a non-Abelian $SU(N_f)_{L}\times SU(N_f)_{R}$ gauge symmetry (dual to the chiral currents) and a bifundamental scalar (dual to the chiral condensate). 
Another interesting direction would be investigating the role of a non-minimal coupling in the phase diagram of QCD in the same line of Refs.~\cite{He:2013qq, Yang:2014bqa, Dudal:2017max, Arefeva:2020byn, Mamani:2020pks, Ballon-Bayona:2020xls}. 
From the gravitational point of view, a natural next step would be the investigation of the quasinormal modes of the black hole solutions found in this work. This would allow us to describe the melting of scalar and tensor glueballs in a non-conformal plasma.

\section*{Acknowledgments}
The authors would like to acknowledge Marco Panero for sharing his results on the thermodynamics of lattice $SU(N_c)$ theories. The authors would like to thank also Jean-Fran\c{c}ois Paquet and the JETSCAPE collaboration for sharing their results on the shear and bulk viscosity, obtained from a model-to-data analysis of the heavy ion collision experimental data. 
The work of A.B-B is partially funded by Conselho Nacional de Desenvolvimento Cient\'\i fico e Tecnol\'ogico (CNPq, Brazil), grants No. 306528/2018-5 and No. 434523/2018-6 and Coordena\c{c}\~ao de Aperfei\c{c}oamento do Pessoal de N\'ivel Superior (CAPES, Brazil), Finance Code 001. L.~A.~H.~M. has financial support from Coordena\c{c}\~ao de Aperfei\c{c}oamento do Pessoal de N\'ivel Superior - Programa Nacional de P\'os-Doutorado (PNPD/CAPES, Brazil). V. T. Z. thanks partial financial support from Coordena\c{c}\~ao de Aperfei\c{c}oamento do Pessoal 
de N\'ivel Superior (CAPES, Brazil), Grant No.~88881.310352/2018-01,
and from Conselho Nacional de Desenvolvimento Cient\'\i fico
e Tecnol\'ogico (CNPq, Brazil), Grant No.~309609/2018-6.

\appendix

\section{Asymptotic analysis}
\label{Sec:Asymptotics}

Here we present some details regarding the asymptotic analysis of the present model, close to the boundary (UV) and close to the horizon (IR). We use these results as ``boundary conditions" to get numerical solutions of the differential equations for the complete model.   

\subsection{Close to the boundary}

The starting point is the asymptotic expansion of the dilaton field \eqref{Eq:DilModelA1} close to the boundary, i.e., in power series for small $z$,
\begin{equation}
\Phi=\phi_{0}\, z^{\Delta_{-}}+ G \,z^{\Delta_{+}}+\cdots,
\end{equation}
\noindent
where $\Delta_-=\epsilon$ and $\Delta_+=4-\epsilon$, and ellipsis denotes higher powers on $z$. Hence, by plugging this expression into the first differential equation of \eqref{EqsBH}, we get the asymptotic expansion for $\zeta_1(z)$ and for the warp factor $A_1(z) = -\ln\zeta_1(z)$,
\noindent
\beqa
\label{asymptotics2}
\zeta_1(z)&=&
\frac{z}{\ell}\left[1+
\frac{2 \Delta_{-}}{9(1+2\Delta_{-})}\phi_{0}^{2} z^{2\Delta_{-}}+
\frac{2 \Delta_{-}\,\Delta_{+}}{45} \phi_{0} G z^4 +
\frac{2\Delta_{+}}{9(1+2\Delta_{+})} G^{2} z^{2\Delta_{+}}
+\dots\right], \cr 
A(z)&=&
-\ln{( z / \ell )}-
\frac{2\Delta_{-}}{9(1+2\Delta_{-})}\phi_{0}^{2} z^{2\Delta_{-}}-
\frac{2\Delta_{-}\Delta_{+}}{45}\phi_{0}G z^4-\dots\,.
\eeqa

The asymptotic form of the horizon function is straightforwardly obtained by plugging \eqref{asymptotics2} into \eqref{EqBlackening}, what gives
\noindent
\begin{equation} \label{eq:f(z)UV}
\begin{aligned}
f(z)&=1-\frac{C_h\,z^4}{4\,\ell^3}\left(1+
\frac{4\,\phi_0^2\,\Delta_{-}}{3(2+\Delta_{-})(1+2\Delta_{-})}
z^{2\Delta_{-}}+\cdots\right).
\end{aligned}
\end{equation}

Similarly, the asymptotic form of the constant $C_h$ is determined by substituting \eqref{asymptotics2} into \eqref{EqIntegConst} and by integrating from $z=z_0$ to $z_h$, with $z_0$ being a UV cutoff, and by neglecting the divergent terms that include $z_0$. The result is
\noindent
\begin{equation}\label{Eq:ChUV}
C_h=\frac{4\ell^3}{z_h^4}\left(1-\frac{4\Delta_{-}\phi_0^2}{3(2+\Delta_-)(1+2\Delta_-)}z_h^{2\Delta_{-}}+\cdots\right), 
\end{equation}
\noindent
where the ellipsis stands for higher power on $z_h$. Notice that the leading term in \eqref{Eq:ChUV} corresponds to the AdS contribution, while the subleading term reveals the deformation introduced by the nontrivial dilaton field.  

One may also write $f(z)$ in terms of $z_h$ by plugging \eqref{Eq:ChUV} into \eqref{eq:f(z)UV}. It is clearly seen that the horizon function reduces to unity in the limit of zero $z$, as expected. 

Let us now calculate the asymptotic expressions of the thermodynamic variables close to the boundary. The asymptotic expansion of the temperature is obtained by plugging \eqref{Eq:ChUV} and \eqref{asymptotics2} into \eqref{EqTemperature},
\noindent
\begin{equation}\label{EqTUV}
\begin{aligned}
T&=\frac{1}{\pi z_h}\left(1+\frac{2\Delta_{-}^2\phi_{0}^2}{3(2+\Delta_-)(1+2\Delta_{-})}z_h^{2\Delta_{-}}+\cdots\right).
\end{aligned}
\end{equation}
\noindent
Again, the leading term is due to the AdS warp factor and the subleading term is the deformation generated by the dilaton field.

It is worth pointing out that in the above analysis the natural independent parameter is the coordinate $z_h$. However, in Thermodynamics one usually uses the temperature as being the fundamental degree of freedom. With this in mind, we invert the asymptotic expression \eqref{EqTUV} to get $z_h$ as a function of the temperature in the form
\begin{equation}\label{Eq:zhTUV}
z_h=\frac{1}{\pi T}\left(1+\frac{2\Delta_{-}^2\phi_0^2}{3(2+\Delta_{-})(1+2\Delta_{-})}\frac{1}{(\pi T)^{2\Delta_{-}}}+\cdots\right).
\end{equation}
With this relation, we may express all thermodynamic variables as a function of the temperature. 

Let us then apply the procedure to the entropy density. By plugging \eqref{Eq:ChUV} and \eqref{asymptotics2} into \eqref{Eq:EntropyDensity} it follows,
\begin{equation}
s=\frac{4\pi\ell^3 \sigma}{z_h^3}\left(1-\frac{2\Delta_{-}\phi_0^2}{3(1+2\Delta_{-})}z_h^{2\Delta_{-}}+\cdots\right)
\end{equation}
Now substituting \eqref{Eq:zhTUV} into the last equation, we get the entropy as a function of the temperature,
\begin{equation}
\frac{s}{\sigma}=4\pi^4\ell^3T^3-\frac{16\pi\ell^3\Delta_{-}\phi_{0}^2}{3(2+\Delta_{-})}(\pi T)^{3-2\Delta_{-}}+\cdots
\end{equation}
Once again the leading term corresponds to the AdS warp factor contribution, which is equivalent to recover conformal symmetry, while the subleading terms correspond to the deformation from such symmetry.

To get an asymptotic expression for the free energy we write the integral representation \eqref{Eq:FreeEnergy} in the form
\begin{equation}\label{eq:FreeUV}
F=\int_{z_h}^{z_{h_{c}}}s(\tilde{z}_h)\left(\frac{dT(\tilde{z}_h)}{d\tilde{z}_h}\right)d\tilde{z}_{h}+\int_{z_{h_c}}^{\infty}s(\tilde{z}_h)\left(\frac{dT(\tilde{z}_h)}{d\tilde{z}_h}\right)d\tilde{z}_{h},
\end{equation}
where $z_{h_c}$ is the value where the temperature reaches its minimal value,  $T(z_{h_c})=T_c$. Equation \eqref{eq:FreeUV} indicates that we may split the free energy in two parts, the first one corresponding to large black holes, $\tilde{z}_h\in [z_h,z_{h_c}]$, and the second one related to small black holes, $\tilde{z}_h\in [z_{h_c},\infty\rangle$. To guarantee the validity of the following analysis we rewrite the free energy of the large black holes in the form
\noindent
\begin{equation}\label{Eq:FreeEnergyBig}
F_{\text{large}}=\int_{z_h}^{z_{h_{*}}}s(\tilde{z}_h)\left(\frac{dT(\tilde{z}_h)}{d\tilde{z}_h}\right)d\tilde{z}_{h}+\int_{z_{h_*}}^{z_{h_{c}}}s(\tilde{z}_h)\left(\frac{dT(\tilde{z}_h)}{d\tilde{z}_h}\right)d\tilde{z}_{h},
\end{equation}
\noindent
where $z_h<z_{h_{*}}<z_{h_{c}}$.

In the large black holes regime, the main contribution is expected to come from the first integral in Eq.~\eqref{Eq:FreeEnergyBig}, and then we may use the asymptotic expressions for the temperature and entropy density to evaluate it. Thus, the result for ${F}_{\text{large}}$ is given by
\begin{equation}
{F}_{\text{large}}=-\frac{\ell^3\sigma}{z_h^4}\left(1-\frac{8\Delta_{-}(1+\Delta_{-}^2)\phi_0^{2}}{3(4-\Delta_{-}^2)(1+2\Delta_{-})}z_h^{2\Delta_{-}}+\cdots\right)+\mathcal{O}(z_{h_*}),
\end{equation}
where $\mathcal{O}(z_{h_*})$ stands for the result of the integral evaluated at $z_{h_*}$, which is a subleading constant term in the limit of zero $z_h$. Hence, as a function of the temperature the free energy becomes
\begin{equation}\label{Eq:FreeEnergyT}
\frac{{F}_{\text{large}}}{\sigma}=-\pi^4\ell^3T^4+\frac{8\ell^3\Delta_{-}\phi_0^{2}}{3(4-\Delta_{-}^2)}(\pi T)^{4-2\Delta_{-}}+\cdots
\end{equation}

Now one may fix the factor $\left(M_{p}\ell\right)^3$ by comparing ${F}$ with the corresponding value obtained by using the Stefan-Boltzmann approximation of pure Yang-Mills, ${F}_{\text{YM}}=-\frac{\pi^2}{45}N_c^{2}\,T^{4}$. Thus, from the leading term of \eqref{Eq:FreeEnergyT} we get
\begin{equation}
\left(M_{p}\ell\right)^3=\frac{1}{45\pi^2},
\end{equation}
which is the same value obtained in Ref.~\cite{Gursoy:2008za}. It is worth pointing out that the subleading term in \eqref{Eq:FreeEnergyT} may be of the same order as the leading term if the value of $\Delta_{-}$ is small enough, such that $T^4\sim T^{4-2\Delta_{-}}$. 

In turn, the asymptotic form of the trace anomaly is given by
\noindent
\begin{equation}
\left\langle T^{\mu}_{\mu}\right\rangle=\frac{16}{3}\frac{\ell^3\phi_0^2\Delta_{-}^2}{(4-\Delta_{-}^2)}(\pi T)^{4-2\Delta_{-}}+\cdots
\end{equation}
\noindent
As it can be seen, the leading term of the trace anomaly goes like $T^{4-2\Delta_{-}}$. In the particular case where the leading term of the dilaton is linear in the UV, i.e., for $\Delta_{-}=1$, this expression reduces to the result obtained in Ref.~\cite{Caselle:2011mn}, with the trace anomaly going like $\langle T_{\mu}^{\mu}\rangle\sim T^2$.

\subsection{Far from the boundary}

So far we have dealt with the asymptotic analysis close to the boundary, and from here on we perform the asymptotic expansion of all relevant quantities far from the boundary, for large $z$. 

Far from the boundary, the dilaton field behaves like the zero temperature asymptotic form
\noindent
\begin{equation}
\Phi=C\,z^2+\cdots
\end{equation}
\noindent
Thus, plugging this expression in the Einstein equation \eqref{EqsBH} we get the asymptotic form for the warp factor, $\zeta_1=e^{-A_1(z)}$,
\begin{equation}\label{Eq:ZetaFunctionSmall}
\begin{split}
\zeta_1=&(\sqrt{C}z)^{-1/2}\text{exp}\left(\frac{2}{3}Cz^2\right)+\cdots.
\end{split}
\end{equation}
\noindent
Now, we may rewrite \eqref{EqIntegConst} in the form
\noindent
\begin{equation}
\int_{0}^{z_h}d\tilde{z}\,\zeta_1^3(\tilde{z})=\int_{0}^{z_{h_c}}d\tilde{z}\,\zeta_1^3(\tilde{z})+\int_{z_{h_c}}^{z_h}d\tilde{z}\,\zeta_1^3(\tilde{z})=\frac{1}{C_h}.
\end{equation}
\noindent
\noindent
Additionally, we may split up the second integral in the intervals $z_{h_{c}}\leq\tilde{z}\leq z_{h_*}$ and $z_{h_{*}}\leq\tilde{z}\leq z_{h}$. Hence, by plugging \eqref{Eq:ZetaFunctionSmall} in \eqref{EqIntegConst} and expanding the result in the region $z_h\gg z_{h_{c}}$, where $z_{h_{c}}$ is the value where the temperature reaches its minimum, the integration constant may be approximated by
\noindent
\begin{equation}\label{Eq:ChIR}
C_h=4\,C^{7/4}z_h^{5/2}e^{-2C z_h^2}\left(1-\frac{5}{8 C z_h^2}+\cdots\right)+\mathcal{O}(z_{h_*}),
\end{equation}
\noindent
where $\mathcal{O}(z_{h_*})$ is the same result evaluated at $z_{h_*}$.

The asymptotic expansion for the temperature close to the singularity is obtained by plugging \eqref{Eq:ChIR} and \eqref{Eq:ZetaFunctionSmall} in expression \eqref{EqTemperature}, what yields 
\noindent
\begin{equation}\label{eq:TIR}
T=\frac{C\,z_h}{\pi}-\frac{5}{8\pi z_h}+\cdots.
\end{equation}
\noindent
This approximate expression confirms the linear behavior observed in our numerical results, as shown by the dashed lines in the left panel of Fig.~\ref{Fig1:zhTFAP}.

In turn, analogously to what we have done in the large black holes regime, we may invert the series \eqref{eq:TIR} in the region of large temperatures. Hence, we get
\begin{equation}\label{Eq:zhTIR}
z_h=\frac{\pi T}{C}+\frac{5}{8\pi T}-\frac{25 C}{64\pi^3 T^3}+\cdots
\end{equation}

The entropy density in this region is given by
\begin{equation}
s=4\pi \sigma e^{-2C z_h^2}(\sqrt{C}z_h)^{3/2}.
\end{equation}
By plugging \eqref{Eq:zhTIR} in the last result we get the entropy density in terms of the temperature,
\begin{equation}
\frac{s}{\sigma}=e^{-\frac{2(\pi T)^2}{C}}\left(\frac{4\pi^{5/2} T^{3/2}}{C^{3/4}}+\frac{15 C^{1/4}\pi^{1/2}}{4 T^{1/2}}+\cdots\right).
\end{equation}
\noindent
The leading term is exponentially suppressed, this behavior can be seen in the right panel of Fig.~\ref{Fig1:zhTFAP} with dashed lines. It is also worth mentioning that the entropy density is always positive. 

Following the same procedure, it is easy to show that the free energy is given by
\noindent
\begin{equation}
\frac{{F}}{\sigma}=e^{-\frac{2(\pi T)^2}{C}}\left(C^{1/4}(\pi T)^{1/2}+\frac{17 C^{5/4}}{16(\pi T)^{3/2}}+\cdots\right)
\end{equation}
\noindent
The exponentially suppressed leading term is also observed in our numerical results, as it is shown by the dashed lines in the left panel of Fig.~\ref{Fig1:TFFAP}.

\end{document}